\pdfoutput=1
\PassOptionsToPackage{table}{xcolor}
\PassOptionsToPackage{pdftex}{graphicx}
\PassOptionsToPackage{pdftex}{xcolor}
\PassOptionsToPackage{pdftex}{pict2e}
\documentclass[10pt,aps,prd,twocolumn,showpacs,superscriptaddress,longbibliography,nofootinbib,floatfix]{revtex4-1}
\usepackage[T1]{fontenc}
\usepackage[utf8]{inputenc}
\DeclareUnicodeCharacter{0130}{\.{I}}
\DeclareUnicodeCharacter{0131}{\i}
\DeclareUnicodeCharacter{00E7}{\c{c}}
\DeclareUnicodeCharacter{00C7}{\c{C}}
\DeclareUnicodeCharacter{011F}{\u{g}}
\DeclareUnicodeCharacter{011E}{\u{G}}
\DeclareUnicodeCharacter{015F}{\c{s}}
\DeclareUnicodeCharacter{015E}{\c{S}}
\DeclareUnicodeCharacter{00F6}{\"{o}}
\DeclareUnicodeCharacter{00D6}{\"{O}}
\DeclareUnicodeCharacter{00FC}{\"{u}}
\DeclareUnicodeCharacter{00DC}{\"{U}}
\DeclareUnicodeCharacter{2013}{--}
\DeclareUnicodeCharacter{2014}{---}
\DeclareUnicodeCharacter{2019}{\'}
\usepackage{xcolor}
\usepackage{graphicx}
\DeclareGraphicsExtensions{.pdf,.png,.jpg,.jpeg,.eps}
\usepackage{mathrsfs,amsmath,amsthm,latexsym,amssymb,amsfonts,cancel,enumerate,txfonts}
\usepackage{diagbox}
\usepackage{multirow}
\usepackage{booktabs}
\usepackage{etoolbox}
\usepackage[colorlinks=true,linkcolor=blue,citecolor=blue,urlcolor=blue]{hyperref}
\usepackage{bookmark}
\usepackage{orcidlink}

\definecolor{navy}{RGB}{0,0,150}

\allowdisplaybreaks

\begin{document}

\title{Particle Dynamics, Shadow and Hawking Sparsity of a Kalb-Ramond Black Hole Coupled to Nonlinear Electrodynamics}

\author{Faizuddin Ahmed\orcidlink{0000-0003-2196-9622}}
\email{faizuddinahmed15@gmail.com}
\affiliation{Department of Physics, The Assam Royal Global University, Guwahati-781035, Assam, India}

\author{Ahmad Al-Badawi\orcidlink{0000-0002-3127-3453}}
\email{ahmadbadawi@ahu.edu.jo}
\affiliation{Department of Physics, Al-Hussein Bin Talal University, 71111,Ma'an, Jordan }

\author{\.{I}zzet Sakall{\i}\orcidlink{0000-0001-7827-9476}}
\email{izzet.sakalli@emu.edu.tr}
\affiliation{Physics Department, Eastern Mediterranean University, Famagusta 99628, North Cyprus via Mersin 10, Turkey}

\begin{abstract}
We study the timelike and null geodesic structure of a static, spherically symmetric black hole sourced by a Kalb--Ramond (KR) field coupled to nonlinear electrodynamics (NED). The geometry is characterized by the mass \(M\), the magnetic monopole charge \(q\), and the Lorentz-violating parameters \((\gamma,\lambda)\). Closed-form expressions are derived for the effective potential, as well as the specific energy and angular momentum of massive particles on circular orbits. We further analyze the photon sphere, black hole shadow, and the Lyapunov exponent associated with unstable null circular geodesics. The latter determines the eikonal quasinormal-mode frequencies through \(\omega_{\rm eik}=(\ell+1/2)\,\Omega_c-i(n+1/2)\,|\lambda_L|.\) The shadow radius is compared with the Event Horizon Telescope (EHT) observations of M87$^\ast$ and Sgr~A$^\ast$, allowing us to identify the viable region in the \((q,\gamma)\) parameter space. Finally, we compute the Hawking temperature, horizon area, and the Gray--Visser sparsity parameter. We demonstrate that the combined effects of the KR field and magnetic monopole charge increase the sparsity parameter from the Schwarzschild value \(16\pi^3 \simeq 496\) to nearly \(1.7\times10^3\). This indicates a significantly sparser Hawking cascade compared to the Schwarzschild case, while the photon ring remains consistent with the EHT \(1\sigma\) observational bounds across most of the physically allowed parameter range.
\end{abstract}
\maketitle

\small

\section{\label{sec:level1}Introduction}

Einstein's field equations admit black hole solutions, and gravitational collapse in General Relativity generically leads to the formation of such compact objects, as established by the celebrated singularity theorems of Penrose and Hawking \cite{Penrose1965,Hawking1970}. Over the past decades, compelling observational evidence has confirmed the existence of black holes across a wide range of masses, from stellar-mass black holes ($M \sim 5$--$20,M_\odot$) to supermassive black holes ($M \sim 10^5$--$10^9,M_\odot$). In the stellar-mass regime, compact objects observed in X-ray binaries possess masses exceeding the theoretical upper bound for neutron stars or quark stars. In the supermassive regime, precise observations of stellar orbits around the compact object at the center of the Milky Way, Sagittarius A* (Sgr A*), indicate that it is too massive and compact to be explained as a cluster of dark stellar remnants or a fermion star.

The observational study of black holes entered a new era with the groundbreaking achievements of the Event Horizon Telescope (EHT) Collaboration, which produced the first horizon-scale images of black hole shadows \cite{EHTL1,EHTL6,EHTL12,EHTL17}. These observations provide a unique opportunity to probe the strong-gravity regime near the event horizon and to test General Relativity as well as possible deviations from it. The theoretical foundation of black hole shadows, however, predates these observations by several decades. Early investigations by Synge \cite{Synge1966} explored the propagation of light in strongly curved spacetimes, while later studies by Bardeen and others established the properties of photon trajectories and shadow formation in Schwarzschild and Kerr geometries \cite{Bardeen1972,Luminet1979}.

A possible violation of local Lorentz symmetry at the particle level arises from the presence of a rank-two antisymmetric tensor field, commonly known as the Kalb-Ramond field \cite{KalbRamond1974}. In string-inspired frameworks, the Kalb-Ramond field naturally appears as a massless excitation of closed strings and is described by an antisymmetric second-rank tensor. Its corresponding rank-three field strength is often interpreted as spacetime torsion in the low-energy effective limit of higher-dimensional string theories, including type IIB theory \cite{Green1987,Altschul2010}. Such torsional formulations have attracted considerable attention due to their potential connection with intrinsic spin, thereby providing a geometric link between string-theoretic excitations and spin-related phenomena in quantum field theory \cite{Hehl1976,Hehl1995}. The theoretical structure of Einstein-Kalb-Ramond (EKR) gravity is constructed from an action that includes both the Einstein--Hilbert term and the antisymmetric KR field, which is non-minimally coupled to spacetime curvature \cite{Lessa2020}. This non-minimal coupling allows for spontaneous Lorentz symmetry breaking when the KR field acquires a non-vanishing vacuum expectation value, thereby modifying the underlying spacetime symmetries. Consequently, the resulting gravitational theory exhibits rich phenomenological features in both asymptotically flat and anti-de Sitter (AdS) spacetimes. These properties make EKR gravity particularly relevant in the context of holography, black hole physics, and broader investigations in quantum gravity phenomenology \cite{Lessa2020,Lessa2021,Atamurotov2022}.

Recent studies on black hole solutions within Kalb--Ramond gravity, in the presence of various exotic matter fields, have explored a wide range of physical properties, including thermodynamics, optical behavior, geodesic structure, perturbations, gravitational lensing, and quasinormal modes. In the static case, black holes with perfect fluid dark matter (PFDM) have been investigated in \cite{Jumaniyozov2025}. Static black holes solution in \cite{Yang2023} and its electrically charged counterpart in \cite{Duan2024} have been studied. Uncharged black holes in the presence of a global monopole have been analyzed in several works \cite{Gullu2022,Belchior2025GlobalMonopole,Fathi2025,Baruah2025}, along with black holes surrounded by a cloud of strings \cite{Silva2025}. Further extensions include charged black holes with PFDM \cite{Faizuddin2026b}, charged black holes in the presence of a cloud of strings \cite{Silva2025,Ahmad2026}, ModMax black holes with a global monopole \cite{Badawi2026}, and dyonic ModMax black holes with a cloud of strings \cite{AhmedSilva2026}. 

A wide variety of nonlinear electrodynamics (NED) theories have been proposed as extensions of Maxwell electrodynamics in order to regularize field singularities and incorporate strong-field quantum corrections. Among the most prominent models are the Born--Infeld (BI) theory \cite{BornInfeld1933,BornInfeld1934}, Euler--Heisenberg (EH) electrodynamics \cite{EulerHeisenberg1936,Schwinger1951}, logarithmic electrodynamics \cite{Soleng1995}. These theories have attracted considerable attention in gravitational physics because they significantly modify black hole geometries, horizon structures, photon trajectories, and consequently black hole shadows. Since then, NED has played a significant role in constructing black hole solutions free from central singularities. Considering nonlinear electromagnetic fields as physical sources of black holes in the context of the General Relativity, it is possible to obtain regular solutions which obey the weak energy condition \cite{Ayon-Beato:2000mjt,Ayon-Beato:1999qin,Ayon-Beato:1998hmi}. 

Several studies have demonstrated that nonlinear electromagnetic corrections can produce observable deviations in photon spheres and shadow radii. In particular, the Born--Infeld model admits analytically tractable magnetically charged black hole solutions whose shadow radius increases with the Born--Infeld nonlinearity parameter \cite{He2022}. Likewise, Euler--Heisenberg and Bronnikov-type NED \cite{Bronnikov2001} black holes have been employed to constrain loop-correction parameters and magnetic charges using Event Horizon Telescope (EHT) observations of the M87$^*$ shadow \cite{Allahyari2020}. The double-logarithmic NED model proposed by G{\"u}ll{\"u} and Mazharimousavi also gives rise to nontrivial black hole solutions with modified thermodynamic and optical properties \cite{Gullu2021}.

A substantial contribution to this field has been made by Kruglov through several nonlinear electromagnetic models and their associated black hole solutions. These include magnetically charged black holes in nonlinear electrodynamics \cite{Kruglov2018}, dyonic black holes with logarithmic electrodynamics \cite{Kruglov2019}, dyonic and magnetized black holes in NED \cite{Kruglov2020}, magnetic black holes with generalized ModMax model of NED \cite{Kruglov2022} as well as black hole solutions arising from two-parameter nonlinear electrodynamics models \cite{Kruglov2015,Kruglov2017}. Kruglov also investigated nonlinear arcsin-electrodynamics and its modified variants, which lead to regularized electromagnetic configurations and novel black hole geometries \cite{Kruglov2016,Kruglov2015b}. In addition, Ma constructed magnetically charged regular black hole solutions within another class of NED models \cite{Ma2015}, while Dymnikova studied regular electrically charged compact objects in nonlinear electrodynamics coupled to gravity \cite{Dymnikova2021}. More recently, nonlinear electrodynamics has also been explored in broader astrophysical and cosmological contexts, including confinement-inspired charged black holes \cite{Mazharimousavi2024}, black holes surrounded by quintessence fields \cite{AlBadawiAhmed2025}, and black holes coupled simultaneously to quintessence and clouds of strings in NED backgrounds \cite{Nascimento2024}. A few more examples of black hole solutions in NED were in \cite{Dymnikova2002,Dymnikova2004,FanWang2016,Rodrigues2016,Balart2014}. Despite the extensive literature on NED-inspired black holes, many analyses remain either perturbative or heavily numerical in nature. Consequently, obtaining analytically manageable black hole solutions and deriving their observable signatures, particularly photon spheres and shadow characteristics, continues to be an important problem in strong-field gravity and black hole phenomenology. 

In this work, we investigate the dynamics of massive particles and shadow of a static, spherically symmetric black hole sourced by NED in the presence of a vacuum expectation value (VEV) of KR field. The KR-VEV acts as a background tensor field that induces spontaneous Lorentz symmetry breaking in spacetime through a non-minimal coupling with the Ricci tensor. We focus on circular motion, in particular innermost stable circular orbits showing the influence of magnetic charge and KR-field parameters. Moreover, we study the photon sphere, black hole shadow, and photon trajectories, which provide important insights into the strong-field regime of gravity. Although the thermodynamic properties of this black hole in anti--de Sitter background was studied in \cite{Singh2026}, its geodesic structure and optical characteristics remain unexplored. This is the gap we address.

The paper is organized as follows. Section~\ref{sec2} presents the background black hole geometry. Section~\ref{sec3} studies massive-particle dynamics and  ISCO. Section~\ref{sec4} treats null geodesics, the photon sphere, the shadow and the corresponding eikonal QNMs, and uses the EHT data to bound the model parameters. Section~\ref{sec5} computes the Hawking temperature and the Gray-Visser sparsity. Section~\ref{sec6} contains the concluding remarks. We use geometrized units, $G=c=\hbar=1$, throughout.

\section{Exact solution of Kalb-Ramond black hole in NED }\label{sec2}

In this section, we investigate black hole solutions in the presence of a KR-field coupled to NED. We begin by considering the action for a self-interacting Kalb-Ramond  field that is non-minimally coupled to gravity, expressed in the form:
\begin{eqnarray}
{\cal {S}}&=&\int d^4x \sqrt{-g}\Big[\frac{R}{2\kappa}-\frac{1}{12}H_{\lambda\mu\nu}H^{\lambda\mu\nu}-V(B_{\mu\nu}B^{\mu\nu}\pm b_{\mu\nu}b^{\mu\nu})\nonumber\\
&+& \frac{1}{2\kappa}(\xi_2B^{\lambda\nu}B^{\nu}_{\mu}+R_{\lambda\mu}+\xi_3B^{\mu\nu}B_{\mu\nu}R)+L(F)\Big],
\label{aa1}
\end{eqnarray}
where $q$ denotes the determinant of the metric tensor, $R$ is the Ricci scalar, $V$ is potential where $B_{\mu\nu}$ represents a self-interacting antisymmetric rank-2 tensor. The parameters $\xi_2$ and $\xi_3$ are the nonminimal coupling constants, while $\kappa = 8\pi G$ is the gravitational coupling constant. The term $\mathcal{L}(F)$ corresponds to the Lagrangian density of the  NLED  source, which is given by
\begin{equation}
{\cal L}(F)=\frac{3}{2sq^2}\left( \frac{\sqrt{2q^2F}}{1+\sqrt{2q^2F}}\right)^{5/2},
\label{aa2}
\end{equation}
with \( s = \frac{q}{2M} \), where \( M \) and \( g \) are free parameters associated with the magnetic monopole's mass and charge, respectively. The energy-momentum tensor (EMT) is obtained by varying Eq.~(\ref{aa2}) with respect to \( A_{\mu} \).

By performing a variation of Eq.~(\ref{aa1}) with respect to the metric \( g_{\mu\nu} \) and the electromagnetic potential \( A_{\mu} \), we obtain the modified Einstein field equations:
\begin{eqnarray}
&&G_{\mu\nu}\equiv R_{\mu\nu}-\frac{1}{2}g_{\mu\nu}R=T_{\mu\nu}^{\rm KR}+T_{\mu\nu}^{\rm NED},\\
&& \nabla_{a}\left(\frac{\partial {{\cal L}}}{\partial F}F^{a b}\right)=0,\\
&&\nabla_{a}(* F^{ab})=0,
\label{aa3}
\end{eqnarray}
where
\begin{widetext}
\begin{eqnarray}
T_{\mu\nu}^{\rm KR}&=& \frac{\xi_2}{\kappa}\left[\frac{1}{2}g_{\mu\nu}B^{\alpha\gamma}B^{\beta}_{\gamma} R_{\alpha\beta}-B^{\alpha}_{\mu} B^{\beta}_{\nu} R_{\alpha\beta}- B^{\alpha\beta}B_{\mu\beta}R_{\nu\beta}-B^{\alpha\beta}B_{\nu\beta}R_{\mu\alpha}\nonumber+\frac{1}{2}D_{\alpha}D_{\mu}(B_{\nu\beta}B^{\alpha\beta})-\frac{1}{2}D^2(B{\alpha}_{\mu}B_{\alpha\nu})-\frac{1}{2}g_{\mu\nu}D_{\alpha}D_{\beta}(B^{\alpha\gamma}B^{\beta}_{\gamma})\right],\label{aa4}\\
T_{\mu\nu}^{\rm NED}&=&2\left[\frac{\partial {{\cal L}}}{\partial F}F_{\mu\sigma}F_{\nu}^{\sigma}-\tilde g_{\mu\nu}{{L(F)}}\right].\label{aa5}
\end{eqnarray}
\end{widetext}

For a static, spherically symmetric solution to the Einstein field equations, the four-dimensional spacetime is described by the following line element:
\begin{equation}
ds^2 = -f(r)dt^2 +\frac{1}{f(r)}dr^2 + r^2\,(d\theta^2+\sin^2 \theta \,d\phi^2).\label{metric}
\end{equation}
The KR-VEV ansatz is
\begin{equation}
b_2 = \tilde{E}(r) \, dt \wedge \lambda \, dr,\label{aa7}    
\end{equation}
with \( b_{tr} = -\tilde{E}(r) \). The squared norm of the field is 
\begin{equation}
b^2 = g^{\mu\nu} g^{\alpha\beta} b_{\mu\alpha} b_{\nu\beta}\label{aa8}    
\end{equation}
is constant relative to (\ref{metric}) provided
\begin{equation}
\tilde E(r) = \sqrt{\tfrac{1}{2}}\,|b|,\label{aa9}
\end{equation}
with $b$ a constant. The function $\tilde E(r)$ defines a static, radial pseudo-electric background
\begin{equation}
\tilde{E}^{\mu} = (0, \tilde{E}(r), 0, 0).\label{aa10}    
\end{equation}

The $rr$-component of the Einstein field equations reads:
 \begin{equation}
 \frac{r^2\lambda}{2}f''(r)+(\lambda+1)rf(r) +f(r)-1=\frac{8M q^2}{(r^3+q^3)^{2}},\label{aa11}
 \end{equation}
where $\lambda \equiv b^2 \xi_2$, and admits \cite{Singh2026}
\begin{equation}
f(r)=1-\frac{2M r^2}{r^3+q^3}+\frac{\gamma}{r^{2/\lambda}}.\label{function}
\end{equation}

The metric (\ref{metric}) with function (\ref{function}) is parameterized by the mass $M$, the magnetic monopole charge $q$, and the Lorentz-violating parameters $\gamma$ and $\lambda$. Setting $q \to 0$ recovers the modified KR black hole reported in \cite{Lessa2020,Atamurotov2022KRBH}. The limit $\gamma\to 0,\,\lambda\to 0$ (equivalently $b^2\to 0$ or $\xi_2\to 0$) reduces (\ref{function}) to the Hayward black hole \cite{Hayward2006}; taking $q,\gamma\to 0$ recovers the Schwarzschild black hole.

For $\gamma\geq 0$, $q=0$ and $\lambda=1$ (with $0<\lambda\leq 2$), Eq.~(\ref{metric}) resembles Reissner--Nordstr\"om \cite{Reissner1916,Nordstrom1918}. The electromagnetic sector differs, however: the pseudo-electric field becomes constant, $E(r)=|b|/\sqrt{2}$, consistent with an asymptotically flat spacetime carrying a spacelike Lorentz-violating background. A localized charge distribution cannot generate a constant electric field, so $\gamma$ is not an electric charge but a Lorentz-violating hair. For $\lambda=2$, the Lorentz-violating source loses its $\gamma$-dependent pieces.

Energy conditions further constrain the parameters. For $\lambda\leq 0$ both the weak and strong conditions are satisfied if $\gamma\leq 0$; in particular, for $\lambda=-1$ they are satisfied only in the presence of a negative cosmological constant. The qualitative change between $\lambda=1$ and $\lambda=-1$ reflects a change in the underlying source structure \cite{Lessa2020,Nandi2023,Atamurotov2022}.

\section{Particle Dynamics}\label{sec3}

In this section, we consider the motion of massive test particles around the geometry of static and spherically symmetric BH in NED within the framework of VEV of the KR field. Using Lagrangian formalism, we derive effective effective potential governs the particles and subsequently study circular motion. In addition, we investigate innermost stable stable circular orbits which gives minimum radius. Related analyses for KR and NLED black holes appear in \cite{AhmedBadawiSakalli2025,AhmedBadawiSakalli2026,Ahmed2026a}.

\subsection{Equation of Motion}

The equation of motion of a massive particle of mass $m$ can be obtained through the Lagrangian density function \cite{Chandrasekhar1983,Wald1984}
\begin{align} 
\mathbb{L}=\frac{1}{2}\,m g_{\mu \nu }\, {u}^{\mu }\, {u}^\nu.\label{bb1} 
\end{align}
where $g_{\mu\nu}$ is the metric tensor and $u^{\mu}=\frac{dx^{\mu}}{d\tau}$ is the four-velocity.

For the considered BH space-time, the Lagrangian density function simplifies as
\begin{equation}
    \mathbb{L}=\frac{m}{2} \left[-f(r) \left(\frac{dt}{d\tau}\right)^2+\frac{1}{f(r)}\,\left(\frac{dr}{d\tau}\right)^2+r^2\,\left(\frac{d\phi}{d\tau}\right)^2\right].\label{bb2}
\end{equation}

The Killing vectors $\xi^\mu_{(t)}\partial_\mu = \partial_t$ and $\xi^\mu_{(\phi)}\partial_\mu = \partial_\phi$ yield conserved energy ${\cal E}=E/m$ and angular momentum ${\cal L}=L/m$ per unit mass. The normalization $g_{\mu\nu}u^\mu u^\nu=-1$, leads to
\begin{align} 
\dot{r}^2= &   {\mathcal {E}}^2+g_{tt}\left( 1+\frac{\mathcal {K}}{r^2}\right) \, \label{bb5}\\
\dot{\theta }= &   \frac{1}{g_{\theta \theta }^2}\Big (\mathcal{K}-\frac{\mathcal{L}^2}{\sin ^2\theta }\Big )\,\label{bb6}\\
\dot{\phi }= &   \frac{\mathcal{L}}{g_{\phi \phi }}\,\label{bb7}\\
\dot{t}= &   -\frac{\mathcal{E}}{g_{tt}}\,\label{bb8}
\end{align}
where $\mathcal{K}$ is the Carter constant.

\subsection{Effective Potential}

The effective potential is a fundamental concept in classical and relativistic mechanics that helps describe the motion of a test particle in a given gravitational or electromagnetic field.

In our further studies, we restricted the motion of test particles in the equatorial plane ($\theta=\pi/2$, $\dot \theta=0$, and $\mathcal{K}=\mathcal{L}^2$), the radial equation simplifies as
\begin{equation}
    \dot{r}^2= {\mathcal {E}}^2-U_{\rm eff},\label{cc1}
\end{equation}
where the effective potential $U_{\rm eff}$ is given by
\begin{equation}
U_{\rm eff}=\left( 1+\frac{\mathcal{L}^2}{r^2}\right)\,f(r).\label{cc2}
\end{equation}

\begin{figure}[ht!]
    \centering
    \includegraphics[width=0.9\linewidth]{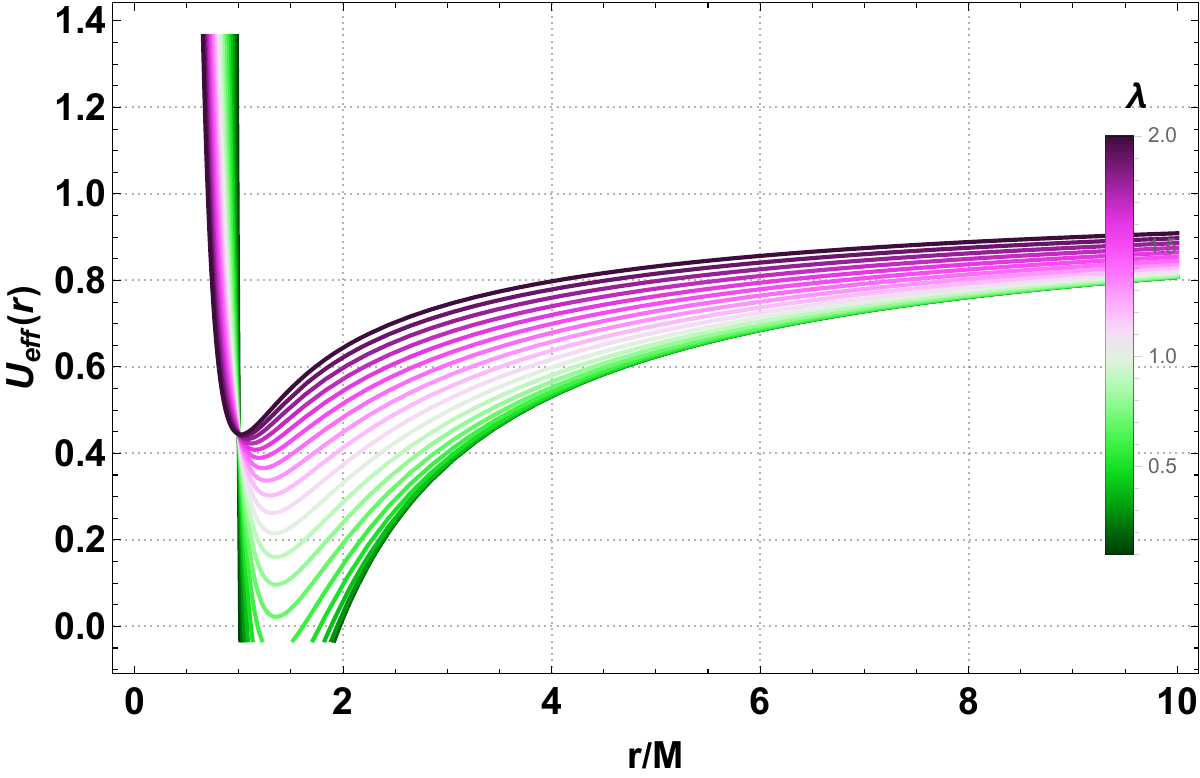}\\
    (i) $q=0.5$\\
    \hfill\\
    \includegraphics[width=0.9\linewidth]{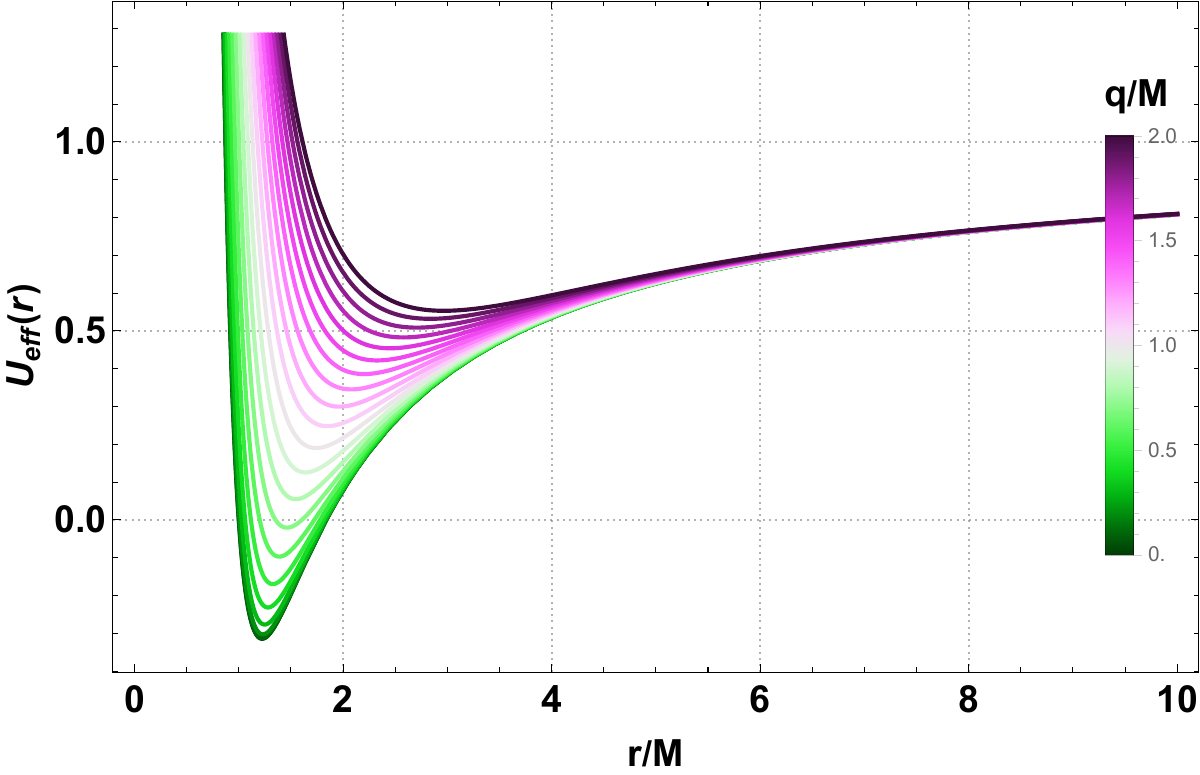}\\
    (ii) $\lambda=0.5$
    \caption{The dependence of effective potential $U_{\rm eff}$
 on radial coordinates $r$ for different values of $\lambda$
 and $q$
 for the massive particle, while keeping $\gamma=1$ fixed.}
    \label{fig:potential}
\end{figure}

Minima of the potential correspond to stable circular orbits, where a small radial perturbation results in bounded oscillations around an equilibrium position. The maxima indicate unstable circular orbits, in which a slight perturbation can cause the particle to fall into the central object or escape to infinity. The steepness and shape of the potential provide insights into the gravitational well and the forces governing orbital motion.

The radial dependence of the effective potential $U_{\rm eff}$ for various values of the parameters $\lambda$ and $q$ for a massive test particle is shown in Figure~\ref{fig:potential}. As can be seen from Fig.~\ref{fig:potential}, an increase in the values of $\lambda$ and $q$ leads to a deeper potential well, indicating a stronger effective gravitational interaction.

\subsection{Circular Orbits}

We now study circular motion of massive test particle around BH using the below conditions 
\begin{equation}
    {\mathcal {E}}^2=U_{\rm eff},\qquad U'_{\rm eff}(r)=0\,,\label{dd1}
\end{equation}
where prime refers to the first-order (partial) derivative w.r. to $r$, corresponds to the minimum values of the effective potential.

By substituting the effective potential in Eq.~(\ref{cc2}) into the circular orbit conditions (\ref{dd1}) and simplifying the resulting expressions, we derive the specific energy and specific angular momentum of a massive particle in circular motion as follows:
\begin{align}
\mathcal{L}^2_{\rm sp}=\frac{\frac{Mr^7-2Mq^3 r^4}{(r^3+q^3)^2}-\frac{\gamma}{\lambda}\, r^{2-\frac{2}{\lambda}}}{1-\frac{2Mr^2}{r^3+q^3}-\frac{Mr^5-2Mq^3 r^2}{(r^3+q^3)^2}+\frac{\gamma\,\left(1+\lambda^{-1}\right)}{r^{2/\lambda}}}\,,\label{angular}\\
\mathcal{E}^2_{\rm sp}=\frac{\left(1-\frac{2M r^2}{r^3+q^3}+\frac{\gamma}{r^{2/\lambda}}\right)^2}{1-\frac{2Mr^2}{r^3+q^3}-\frac{Mr^5-2Mq^3 r^2}{(r^3+q^3)^2}+\frac{\gamma\,\left(1+\lambda^{-1}\right)}{r^{2/\lambda}}}\,.\label{energy}
\end{align}

\begin{figure}[ht!]
    \centering
    \includegraphics[width=0.9\linewidth]{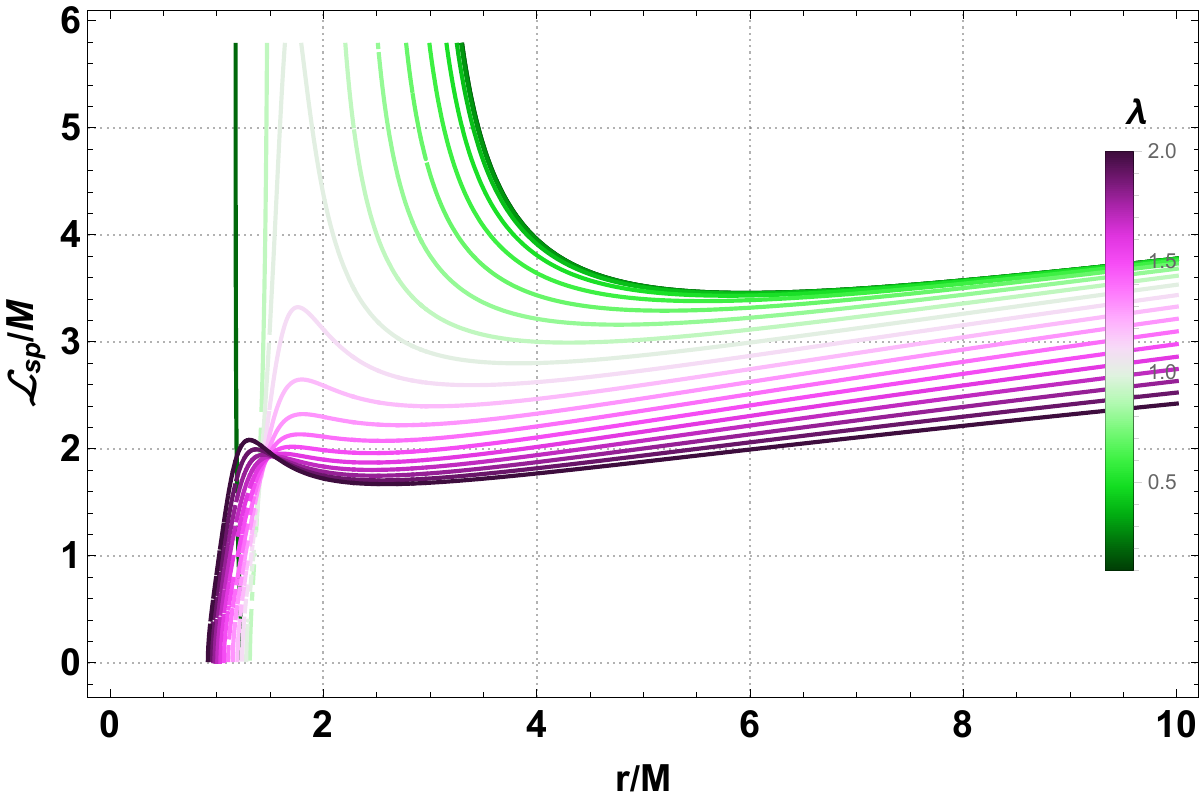}\\
    (i) $q/M=0.5$\\
    \hfill\\
    \includegraphics[width=0.9\linewidth]{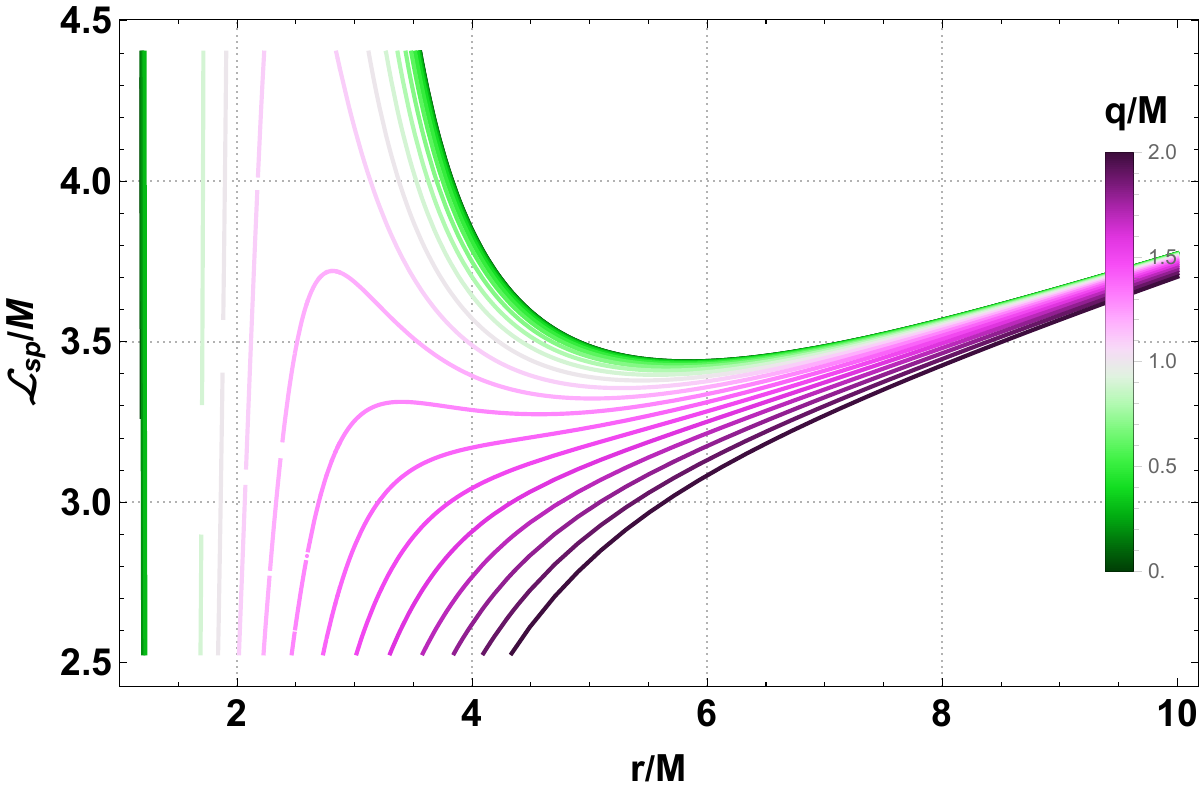}\\
    (ii) $\lambda=0.5$
    \caption{The dependence of the specific angular momentum $\mathcal{L}_{\rm sp}$ on radial coordinates $r$ for different values of $\lambda$ and $q$ for the massive particle, while keeping $\gamma=1=M$ fixed.}
    \label{fig:momentum}
\end{figure}

\begin{figure}[ht!]
    \centering
    \includegraphics[width=0.9\linewidth]{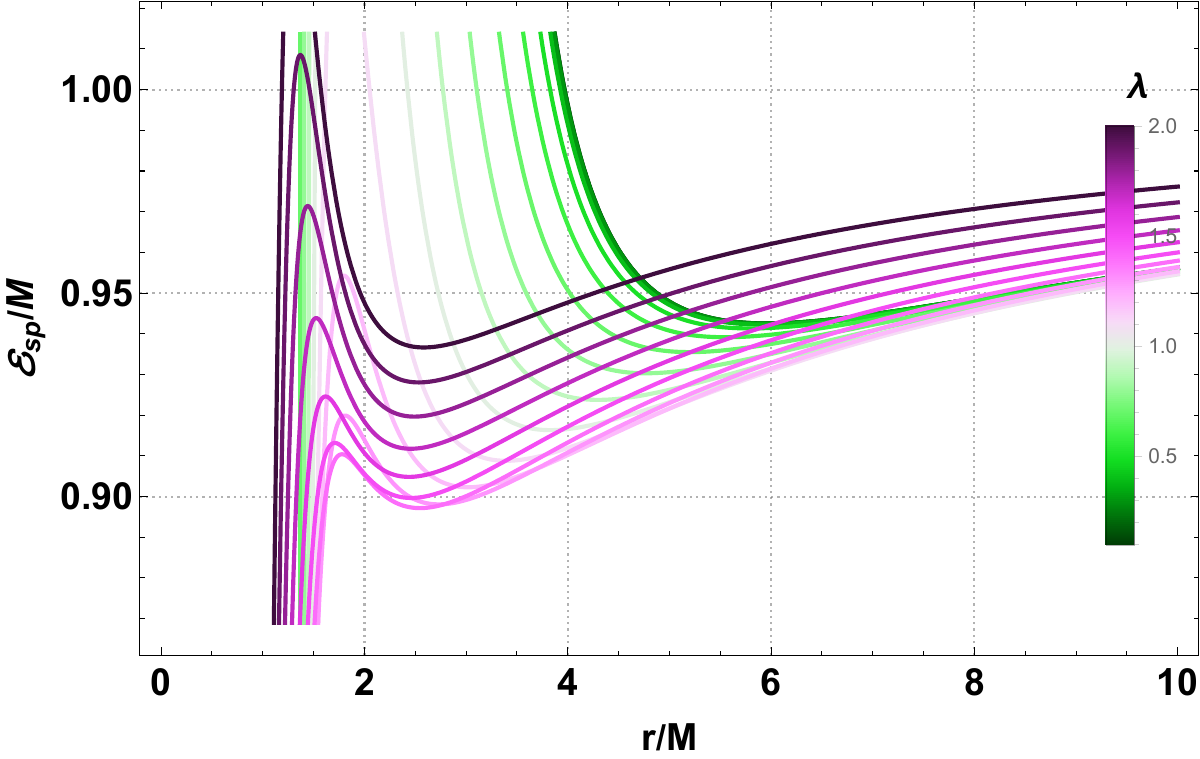}\\
     (i) $q=0.5$\\
     \hfill\\
    \includegraphics[width=0.9\linewidth]{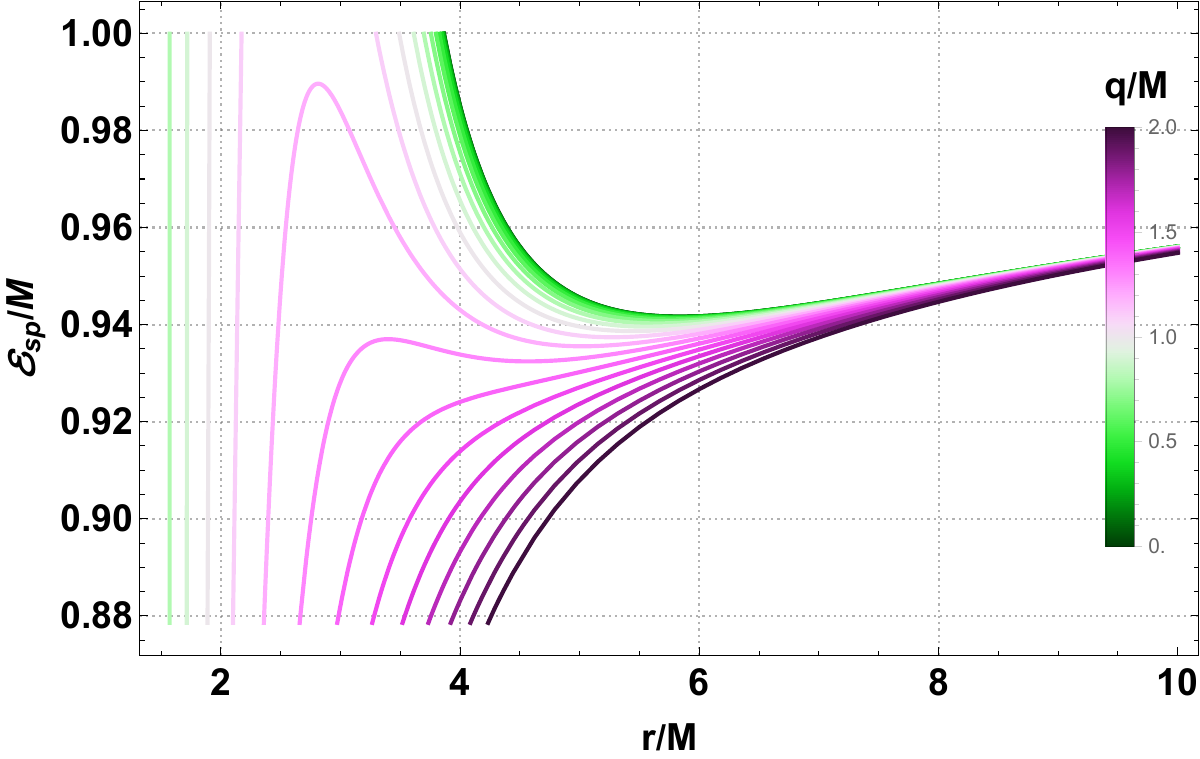}\\
   (ii) $\lambda=0.5$
    \caption{The dependence of the specific energy  $\mathcal{E}_{\rm sp}$ on radial coordinates $r$ for different values of $\lambda$ and $q$ for the massive particle, while keeping $\gamma=1=M$ fixed.}
    \label{fig:energy}
\end{figure}

The radial dependence of the specific angular momentum $\mathcal{L}_{\rm sp}$ and specific energy $\mathcal{E}_{\rm sp}$ for various values of the parameters $\lambda$ and $q$ for a massive test particle is shown in Figure~\ref{fig:momentum} and Figure~\ref{fig:energy}, respectively. As seen from both Figures, increasing the values of $\lambda$ and $q$ leads to a reduction in the magnitude of both $\mathcal{L}_{\rm sp}$ and $\mathcal{E}_{\rm sp}$. 

This behavior indicates that the circular motion becomes energetically less demanding as these parameters increase, implying a modification of the effective gravitational potential in the strong-field region. In particular, the innermost stable circular orbit (ISCO) shifts inward, resulting in smaller ISCO radii that require lower angular momentum and energy for a particle to maintain stable circular motion. Consequently, for smaller values of $\lambda$ and $q$, test particles remain more tightly bound to the black hole, whereas larger values weaken the binding and reduce the orbital requirements.

\subsection{Innermost stable circular orbits: ISCO}

The innermost stable circular orbit (ISCO) represents the smallest radius at which a test particle can maintain a stable circular trajectory. Beyond the ISCO, circular orbits become unstable, leading to a transition to plunge into the BH or move outward.

Stable circular orbits occur at radius $r=r_{\rm min}$, where the particles' minimum energy and angular momentum correspond to circular orbits. For the innermost stable circular orbits (ISCOs), the following conditions must be satisfied \cite{Chandrasekhar1983,Wald1984}
\begin{equation}
     {\mathcal {E}}^2=U_{\rm eff},\quad U'_{\rm eff}(r)=0,\quad U''_{\rm eff} \geq 0.\label{ee1}
\end{equation}

For a marginally stable circular orbits, where the condition $U''_{\rm eff}(r)=0$ is satisfied, we find the following relations:
\begin{equation}
    f(r)\,f''(r)-2\bigl(f'(r)\bigr)^2+\frac{3\,f(r)\,f'(r)}{r}=0. \label{ee2}
\end{equation}
The highly nonlinear nature of Eq.~(\ref{ee2}) for the given metric function $f(r)$ precludes a general exact analytical solution for ISCO radius. Therefore, in the present work, we resort to a numerical determination of the ISCO radius $r=r_{\rm ISCO}$ for physically admissible values of $\lambda$ and $q$. Restricting our analysis to the interval  \(0 \leq \lambda \leq 2, \, \gamma > 0.\), we select the representative Lorentz-violating parameter value $\lambda=0.5$, and solve the ISCO equation numerically for each case. (see Table \ref{tab:isco-radius}).

\begin{table}[ht!]
\centering
\caption{Numerical values of $ISCO$ for different $q$ and $\gamma$. Here, we set $M=1,\,\lambda=0.5$.}
\begin{tabular}{|c|c|c|c|c|c|}
\hline
$q (\downarrow)$ $ \gamma (\rightarrow )$ & 0.1 & 0.2 &0.3 & 0.4 & 0.5 \\ \hline 
 0.0 &5.9848 & 5.9699 &5.9534  & 5.9375& 5.9214 \\
\hline
0.2 & 5.9824 & 5.9666 &5.9508  & 5.9349 &5.9188 \\
\hline
0.4 & 5.9647 & 5.9490 &5.9330  & 5.9169 & 5.9005\\
\hline
0.6 & 5.9164 & 5.9001& 5.8835 & 5.8668 &5.8499  \\
\hline
0.8 &5.8180  & 5.8005 & 5.7818 &5.7636  &5.7451 \\
\hline
\end{tabular}
\label{tab:isco-radius}
\end{table}

\begin{figure}[ht!]
    \centering
    \includegraphics[width=0.9\linewidth]{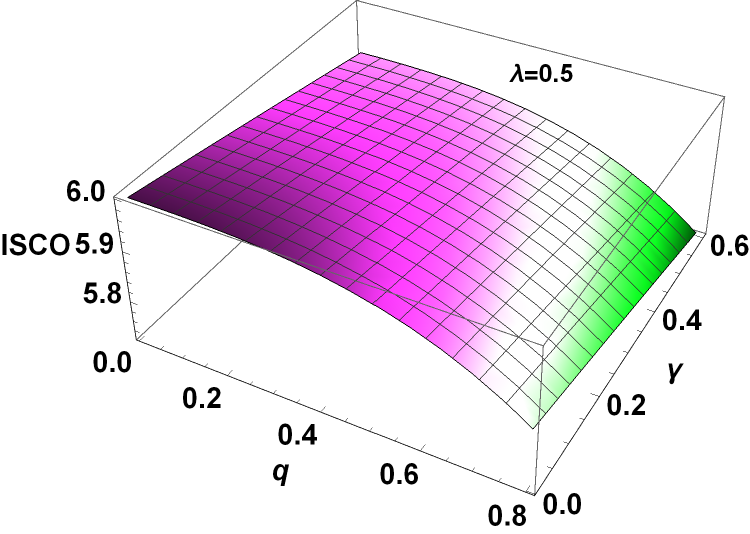}
    \caption{Behavior of the ISCO radius for numerous values of BH parameters $q$ and $\gamma$.}
    \label{fig:isco}
\end{figure}

Figure~\ref{fig:isco} illustrates the behavior of the ISCO radius, $r_{\rm ISCO}$, as a function of $q$ and $\gamma$. The Schwarzschild limit, $r_{\rm ISCO}=6M$, is correctly recovered for $q=\gamma=0$. As $q$ increases, the ISCO radius decreases smoothly, indicating that stable circular orbits move closer to the black hole. The $\gamma$ further modulates this behavior, where larger values of $\gamma$ shift the ISCO curves downward, corresponding to smaller orbital radii and reflecting a weaker effective gravitational binding. Nevertheless, the effect of $q$ is more pronounced than that of $\gamma$.

\subsection{Energy efficiency and orbital frequency}

Now, we analyze the dependencies of energy ($\mathcal{E}|_{\rm ISCO}$) and the angular momentum $(\mathcal{L}|_{\rm ISCO})$ of the particles at ISCO, as well as the energy efficiency $\eta=1-\mathcal{E}|_{\rm ISCO}$. At radius $r=r_{\rm ISCO}$, the angular momentum and energy are given by
\begin{align}
\mathcal{L}^2|_{\rm ISCO}=\frac{r^3_{\rm ISCO}\,f'(r_{\rm ISCO})}{2\,f(r_{\rm ISCO})-r_{\rm ISCO}\,f'(r_{\rm ISCO})}\,,\label{ff1}\\
\mathcal{E}^2|_{\rm ISCO}=\frac{2 f^2(r_{\rm ISCO})}{2\,f(r_{\rm ISCO})-r_{\rm ISCO}\,f'(r_{\rm ISCO})}\,.\label{ff2}
\end{align}
Therefore, the accretion efficiency is obtained as, 
\begin{equation}
\eta=1-\mathcal{E}|_{\rm ISCO}=1-\sqrt{\frac{2 f^2(r_{\rm ISCO})}{2\,f(r_{\rm ISCO})-r_{\rm ISCO}\,f'(r_{\rm ISCO})}}\,.\label{ff3}
\end{equation}

The orbital velocity of massive particles in circular paths of fixed radii is defined by
\begin{align}
    \Omega_{\phi} \equiv \frac{d\phi}{dt}&=\frac{\dot \phi}{\dot t}=\sqrt{\frac{f'(r)}{2 r}}\nonumber\\
    &=\sqrt{\frac{Mr^3-2Mq^3}{(r^3+q^3)^2}-\frac{\gamma}{\lambda}\, r^{-2/\lambda-2}},\label{velocity}
\end{align}
where we have used the relation $\frac{\mathcal{L}}{\mathcal{E}}=\frac{r}{f(r)}\sqrt{\frac{r f'}{2}}$ from Eqs.~(\ref{angular})-(\ref{energy}).

\begin{figure}[ht!]
    \centering
    \includegraphics[width=0.9\linewidth]{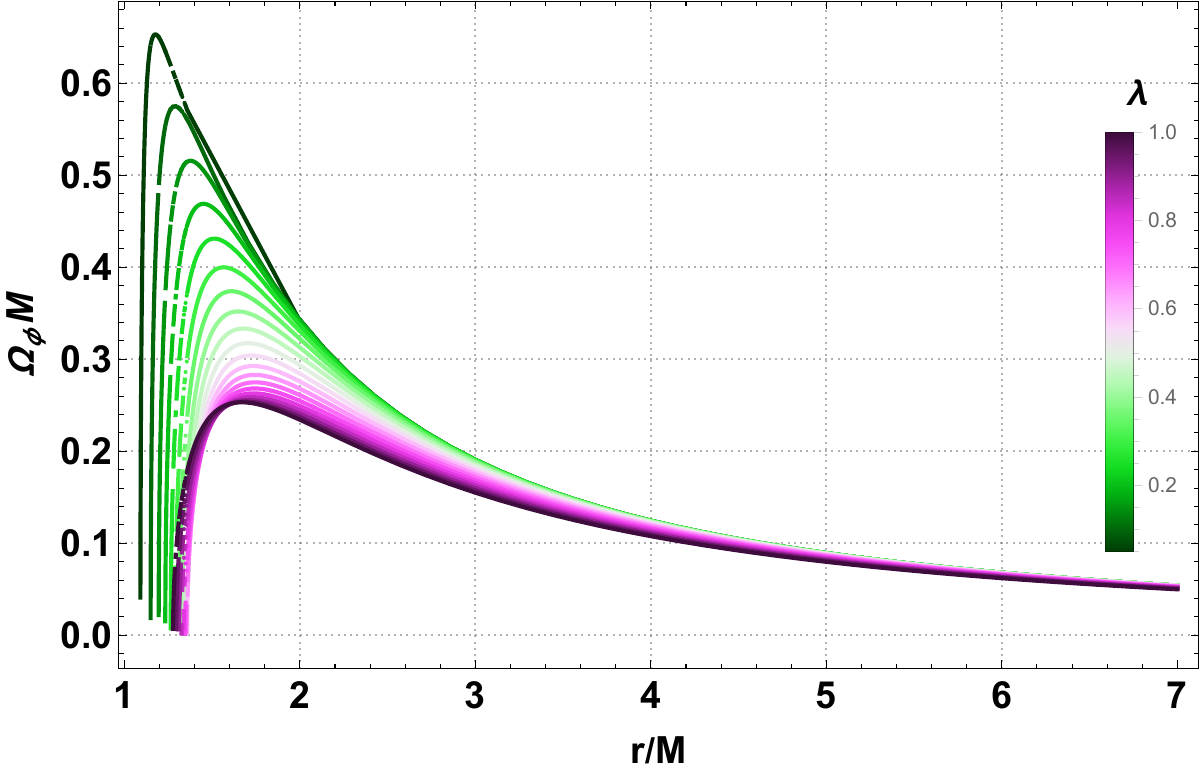}\\
    (i) $q=0.5$\\ 
    \hfill\\
    \includegraphics[width=0.9\linewidth]{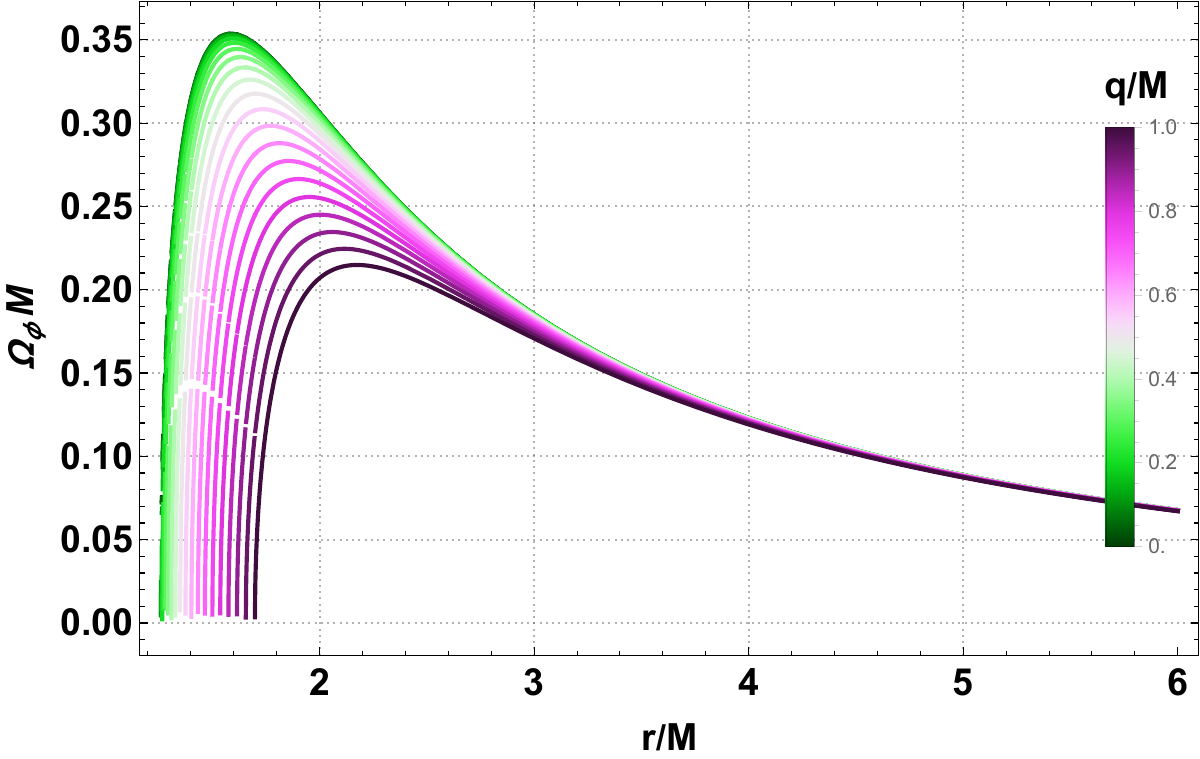}\\
    (ii) $\lambda=0.5$
    \caption{The dependence of the orbital velocity $\Omega_{\phi}$ on radial coordinates $r$ for different values of $\lambda$ and $q$ for the massive particle, while keeping $\gamma=1=M$ fixed.}
    \label{fig:velocity}
\end{figure}

The radial dependence of $\Omega_{\phi}$ for various values of $\lambda$ and $q$ is shown in Figure~\ref{fig:velocity}. As can be seen from the figure, an increase in $\lambda$ leads to a reduction in the orbital velocity in circular orbits. In contrast, higher values of $q$ result in an enhancement of $\Omega_{\phi}$. These opposite trends indicate that these two parameters influence the dynamical behavior of test particles in distinctly different ways, with $\lambda$ effectively weakening while $q$ strengthens the rotational motion in the equatorial plane.

From the above analysis of particle dynamics, it is evident that the magnetic monopole charge $q$ and the KR field parameters $(\gamma,\lambda)$ play a crucial role in modifying the effective potential governing particle motion. These parameters, in turn, lead to significant deviations in the specific energy and specific angular momentum in circular orbits. As a result, the characteristics of ISCOs are notably altered, along with the corresponding orbital velocity profiles. This demonstrates that both the magnetic charge and the KR field contributions have a direct impact on the stability and kinematic properties of particle trajectories in the black hole spacetime.

\section{Null Geodesics: Photon Dynamics}\label{sec4}

The motion of photons around a black hole is governed by null geodesics of the underlying spacetime geometry. Since photons are massless particles, their trajectories satisfy the null condition $ds^{2}=0$, which leads to a set of coupled differential equations describing the propagation of light in the curved spacetime background. The analysis of null geodesics plays a fundamental role in understanding several important optical phenomena associated with black holes, including gravitational lensing, photon capture, black hole shadows, and the formation of photon spheres \cite{Chandrasekhar1983,Wald1984}. In particular, unstable circular photon orbits \cite{Cardoso2009} determine the boundary of the black hole shadow observed by a distant observer. In nonlinear electrodynamics-inspired black hole spacetimes, modifications to the metric function can significantly alter the effective potential experienced by photons, thereby affecting the radius of the photon sphere, and the size and shape of the shadow. Consequently, the study of null geodesics offers an important observational window into the strong gravity regime and the underlying nonlinear electromagnetic corrections encoded in the black hole geometry.

The Lagrangian function for photons around a static spherically symmetric BH metric (\ref{metric}) is
\begin{equation}
    2\mathbb{L}=-f(r) \dot t^2+\frac{\dot r^2}{f(r)}+r^2 (\dot \theta^2+\sin^2 \theta \dot \phi^2),\label{gg1}
\end{equation}
Considering the photon motion in the equatorial plane, $\theta=\pi/2$, the equations of motion are 
\begin{align}
    \dot t=& \frac{\mathrm{E}}{f(r)},\,\label{gg2}\\
    \dot \phi=& \frac{\mathrm{L}}{r^2},\,\label{gg3}\\
    \dot r^2=& \mathrm{E}^2-V_{\rm eff},\,\label{gg4}
\end{align}
where the effective potential $V_{\rm eff}$ is given by
\begin{equation}
    V_{\rm eff}=\frac{\mathrm{L}^2}{r^2}\,f(r).\,\label{gg5}
\end{equation}

\begin{figure}[ht!]
    \centering
    \includegraphics[width=0.9\linewidth]{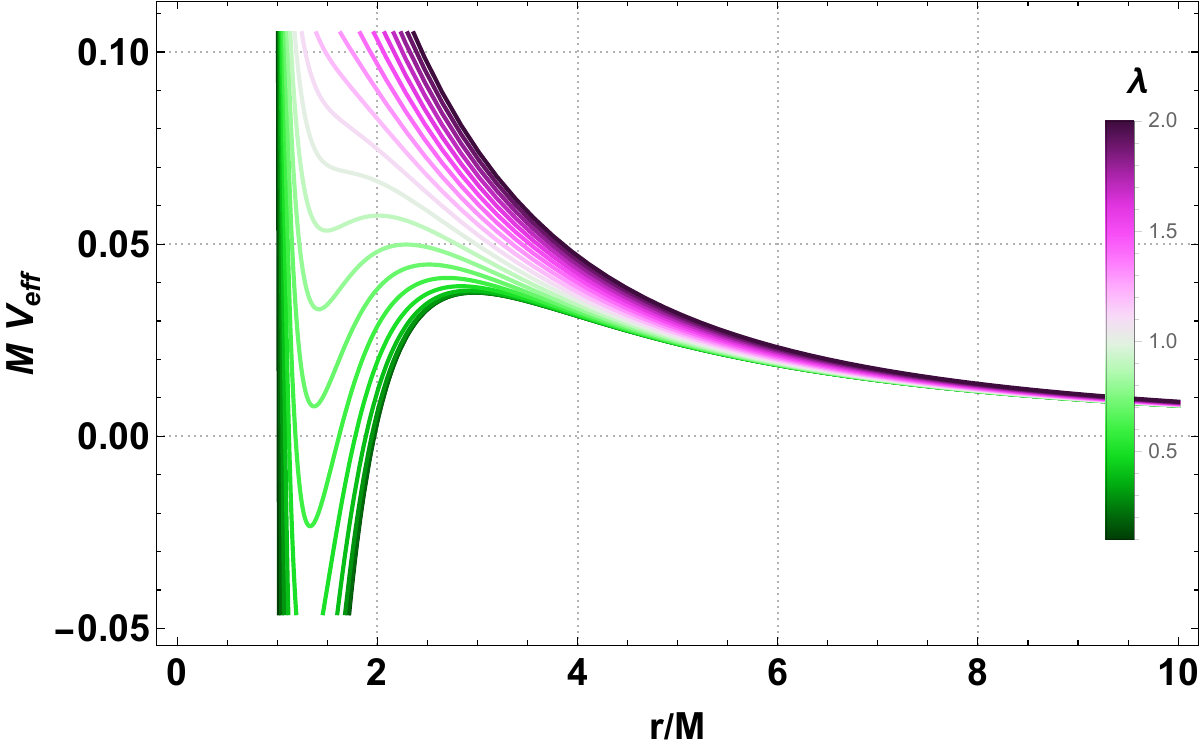}\\
    (i) $q=0.5$ \\
    \includegraphics[width=0.9\linewidth]{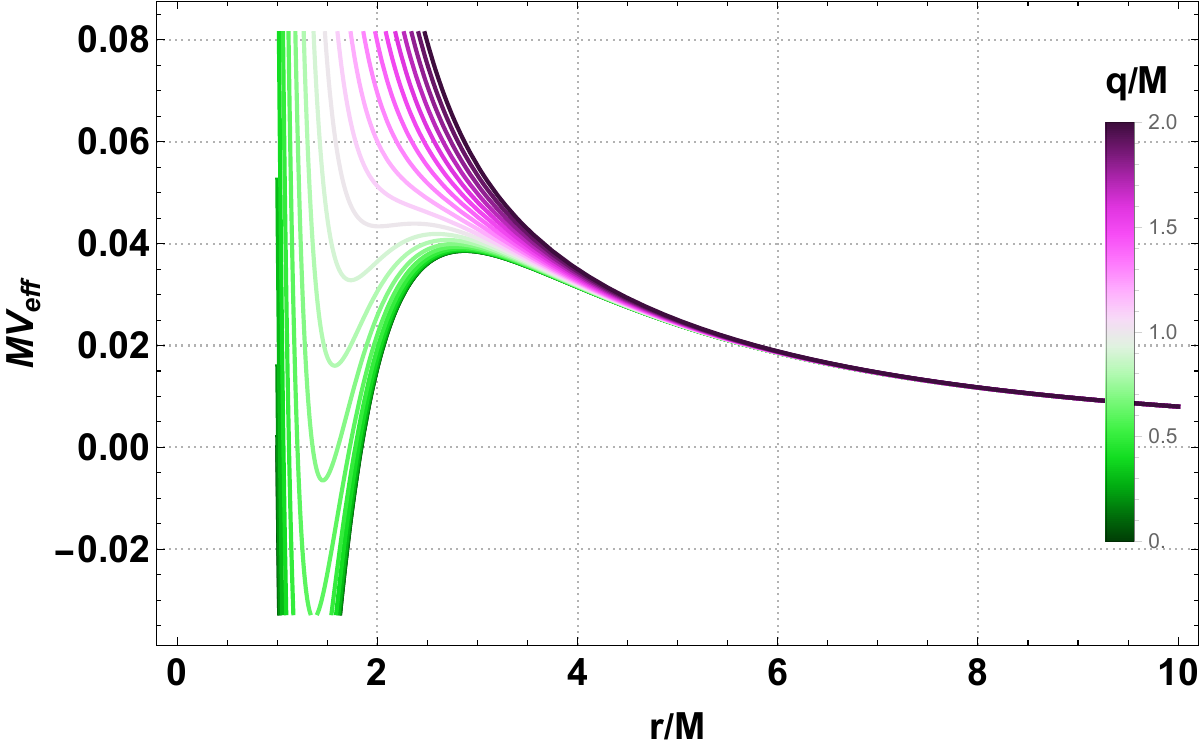}\\
    (ii) $\lambda=0.5$
    \caption{The dependence of the effective potential $V_{\rm eff}$ for null geodesics on radial coordinates $r$ for different values of $\lambda$ and $q$ for photons, while keeping $\gamma=1=M$ fixed.}
    \label{fig:potential-null}
\end{figure}

The sign of $V''_{\rm eff}$ classifies the stability of photon circular orbits: $V''_{\rm eff}>0$ for stable, $V''_{\rm eff}<0$ for unstable ones \cite{Cardoso2009}.

Figure~\ref{fig:potential-null} shows the radial behavior of the effective potential \(V_{\rm eff}\) for different values of \(\lambda\) and \(q\). Increasing these parameters enhances the effective potential, indicating stronger spacetime curvature effects on null geodesics. The resulting modification of the potential barrier affects the location and instability of circular photon orbits, thereby influencing the photon sphere radius, black hole shadow size, and strong-field light deflection.

\subsection{Photon sphere and Shadow}

Imposing $\dot r=0$ and $V'_{\rm eff}(r)=0$ yields
\begin{equation}
    \mathrm{E}^2=\frac{\mathrm{L}^2}{r^2}\,f(r)\Longrightarrow \beta_c=\frac{\mathrm{L}}{\mathrm{E}}\Big{|}_{r=r_c}=\frac{r}{\sqrt{f(r)}}\Big{|}_{r=r_c}.\label{critical}
\end{equation}
And
\begin{equation}
    \frac{\partial}{\partial r}\left(\frac{f(r)}{r^2}\right)=0\Longrightarrow 2\,f(r)-r\,f'(r)=0.\, \label{photon1}
\end{equation}
which determines the photon-sphere radius $r=r_s$. 

We now investigate the black hole shadow and the influence of the NED and KR fields on its properties. Existing theoretical and observational constraints restrict the Lorentz symmetry breaking parameters to \(0 \leq \lambda \leq 2\) and \(\gamma>0\), consistent with the weak energy condition \cite{Lessa2020}. In addition, EHT observations constrain \(\lambda \lesssim 1.2\,(1\sigma)\) and \(\lambda \lesssim 1.5\,(2\sigma)\) for \(\gamma=1/2\) \cite{Vagnozzi2023}. Within these bounds, the metric remains asymptotically flat \((f(r)\to1 \text{ as } r\to\infty)\), implying that the shadow radius is given by the critical impact parameter evaluated at \(r=r_s\) \cite{Volker2022}. Using Eq.~(\ref{critical}), the shadow radius is obtained as follows:
\begin{equation}
R_{\rm sh}=\frac{r_s}{\sqrt{f(r_s)}}=\frac{r_s}{\sqrt{1-\dfrac{2 M r_s^2}{r_s^3+q^3}+\dfrac{\gamma}{r_s^{2/\lambda}}}}.\label{hh6}
\end{equation} 
Eq.~(\ref{photon1}) becomes
\begin{equation}
    2-\frac{6Mr^5}{(q^3+r^3)^2}+\frac{2\gamma(1+\lambda)}{\lambda \,r^{2/\lambda}}=0.\label{photon2}
\end{equation}
Equation~(\ref{photon2}) does not admit a closed form; it is solved numerically for $\lambda=0.5$ and the resulting $r_s$, $R_{\rm sh}$ values appear in Table~\ref{tab:photon-radius}.

\begin{table*}[ht!]
\centering
\renewcommand{\arraystretch}{1.45}
\setlength{\tabcolsep}{6pt}
\begin{tabular}{|c|cc|cc|cc|cc|cc|}
\hline
$ \gamma (\rightarrow )$ & \multicolumn{2}{|c|} {0.1} & \multicolumn{2}{|c|} {0.2}& \multicolumn{2}{|c|} {0.3} & \multicolumn{2}{|c|} {0.4} & \multicolumn{2}{|c|} {0.5} \\ \hline 
 $q (\downarrow)$ & \multicolumn{2}{c|}{$r_{\rm s}$\hspace{0.6 cm} $R_{\rm sh}$} & \multicolumn{2}{c|}{$r_{\rm s}$\hspace{0.6 cm} $R_{\rm sh}$} & \multicolumn{2}{c|}{$r_{\rm s}$\hspace{0.6 cm} $R_{\rm sh}$} & \multicolumn{2}{c|}{$r_{\rm s}$ \hspace{0.6 cm}$R_{\rm sh}$}& \multicolumn{2}{c|}{$r_{\rm s}$ \hspace{0.6 cm}$R_{\rm sh}$} \\ \hline
0.0 & 2.9887 & 5.18645 & 2.9771 &5.17658 & 2.9655&5.16653 & 2.9534& 5.1563& 2.9410&5.14587 \\
\hline
0.2 & 2.9869 & 5.18488 & 2.9752 & 5.17498 &2.9636 &5.1649 &2.9515 & 5.15464&2.9390&5.14418 \\
\hline
0.4 & 2.9738 &5.17381  & 2.9622 & 5.16371 &2.9500 &5.15343 & 2.9376& 5.14295&2.9248&5.13226 \\
\hline
0.6 & 2.9377 & 5.14288& 2.9249 & 5.13219 & 2.9117&5.12128 &2.8982 &5.11014 &2.8842&5.09876 \\
\hline
0.8 & 2.8600 &5.07831  & 2.8448 &5.0662  &2.8290 &5.05377 &2.8126 & 5.04101&2.7955&5.02789 \\
\hline
\end{tabular}
\caption{Photon-sphere radius $r_s$ and shadow radius $R_{\rm sh}$ for different $q$ and $\gamma$, with $M=1,\,\lambda=0.5$.}
\label{tab:photon-radius}
\end{table*}

\begin{figure}[ht!]
    \centering
    \includegraphics[width=0.9\linewidth]{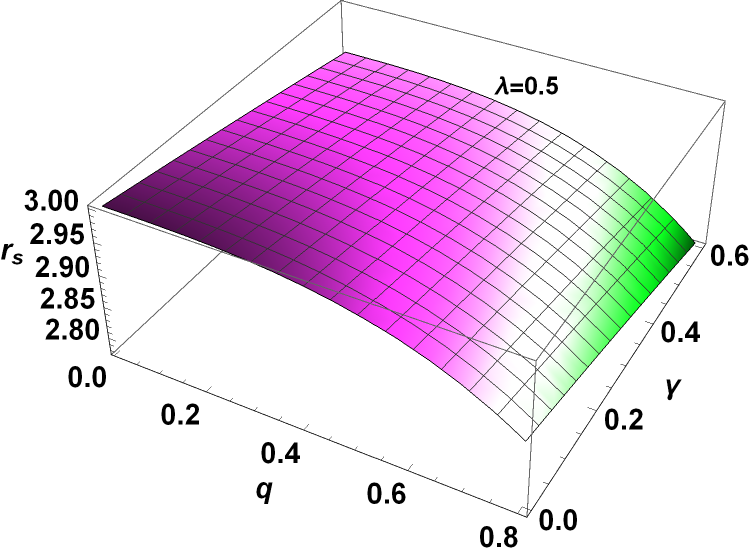}\\
    \includegraphics[width=0.9\linewidth]{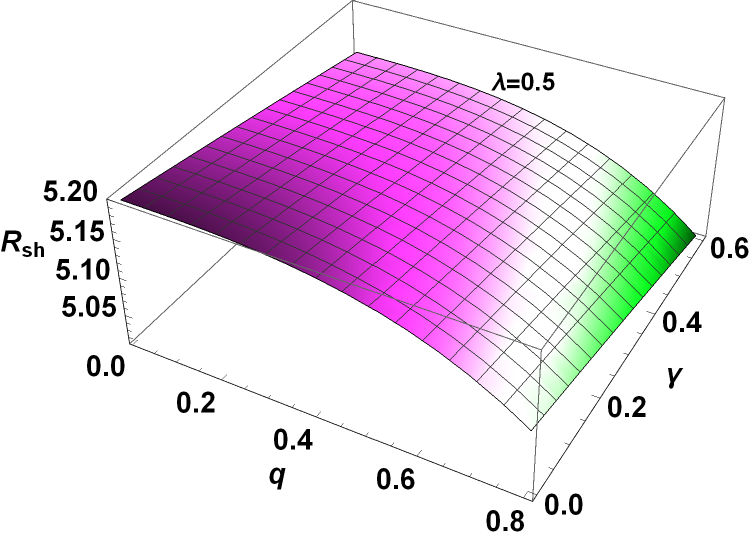}
    \caption{Behavior of the photon sphere $r_s$ and shadow radius $R_{s}/M$ for numerous values of BH parameters $q$ and $\gamma$. Here $M=1,\,\lambda=0.5$.}
    \label{fig:shadow}
\end{figure}

Figure~\ref{fig:shadow} confirms that $r_s$ and $R_{\rm sh}$ both shrink as $q$ grows, with $\gamma$ providing a milder downward shift. The Schwarzschild values $r_s=3M$ and $R_{\rm sh}=3\sqrt{3}\,M\simeq 5.196\,M$ are recovered at $q=\gamma=0$.

\subsection[EHT compatibility for M87* and Sgr A*]{EHT compatibility for M87$^\ast$ and Sgr~A$^*$}\label{isec4-EHT}

The 1$\sigma$ EHT bands on the shadow radius for M87$^\ast$ and Sgr~A$^\ast$ read \cite{Vagnozzi2023,WOS:000571990600005,EHTL17}
\begin{equation}
4.31\leq R_{\rm sh}^{\rm M87^\ast}/M\leq 6.08,\qquad 4.55\leq R_{\rm sh}^{\rm SgrA^\ast}/M\leq 5.22.\label{EHT_bounds}
\end{equation}
The Sgr~A$^\ast$ band is the more restrictive of the two. For every $(q,\gamma)$ pair in Table~\ref{tab:photon-radius}, $R_{\rm sh}/M$ falls inside both bands; the model is therefore compatible with present EHT data over the full range scanned. Table~\ref{tab:EHT-window} reports the upper bound on $\lambda$ obtained by holding $q=0$ and pushing $\lambda$ until $R_{\rm sh}$ leaves the M87$^\ast$ or Sgr~A$^\ast$ window.

\begin{table}[ht!]
\centering
\renewcommand{\arraystretch}{1.45}
\setlength{\tabcolsep}{10pt}
\begin{tabular}{|c|c|c|}
\hline
$\gamma$ & $\lambda_{\rm max}$ (M87$^\ast$) & $\lambda_{\rm max}$ (Sgr~A$^\ast$) \\ \hline
0.05     & $<2$    & $<2$    \\ \hline
0.10     & $<2$    & $<2$    \\ \hline
0.20     & $<2$    & $<2$    \\ \hline
0.30     & $<2$    & 1.673   \\ \hline
0.40     & 1.700   & 1.334   \\ \hline
0.50     & 1.400   & 1.151   \\ \hline
\end{tabular}
\caption{Allowed upper bound $\lambda_{\rm max}$ for which $R_{\rm sh}/M$ remains inside the EHT 1$\sigma$ band, at $q=0$, $M=1$. Entries marked ``$<2$'' indicate that the entire scanned range $\lambda\in[0.05,2]$ is allowed.}
\label{tab:EHT-window}
\end{table}

The Sgr~A$^\ast$ data tighten the bound on $\lambda$ noticeably faster than the M87$^\ast$ data. For $\gamma\gtrsim 0.4$ the allowed $\lambda$ window narrows; small $\gamma$ leaves the model essentially unconstrained.

\subsection{Effective radial force and orbit Stability}

The effective radial force experienced is the negative gradient of potential given by
\begin{align}
    \mathcal{F}=-\frac{1}{2}\frac{\partial V_{\rm eff}}{\partial r}=\frac{\mathrm{L}^2}{r^3}\left[1-\frac{2Mr^2}{r^3+q^3}-\frac{Mr^5-2Mq^3 r^2}{(r^3+q^3)^2}+\frac{\gamma\,\left(1+\lambda^{-1}\right)}{r^{2/\lambda}}\right]\,.\label{hh1}
\end{align}

\begin{figure*}[tbhp]
    \centering
    \includegraphics[width=0.4\linewidth]{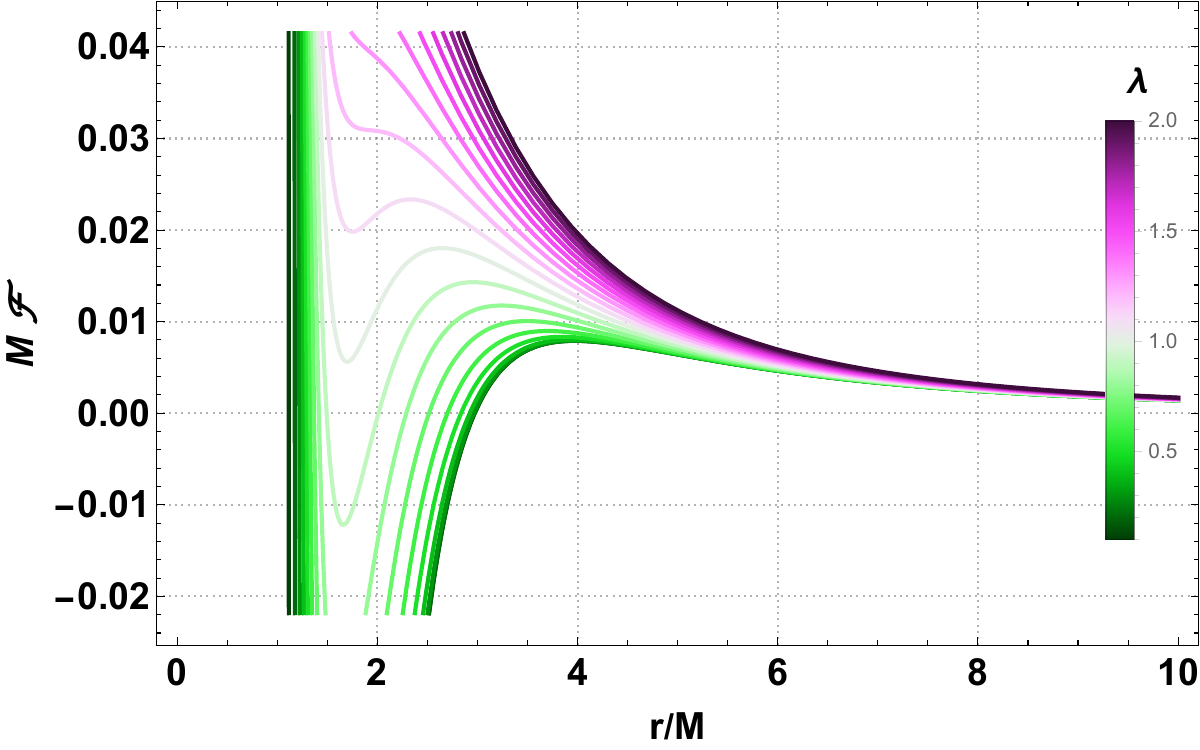}\qquad
    \includegraphics[width=0.4\linewidth]{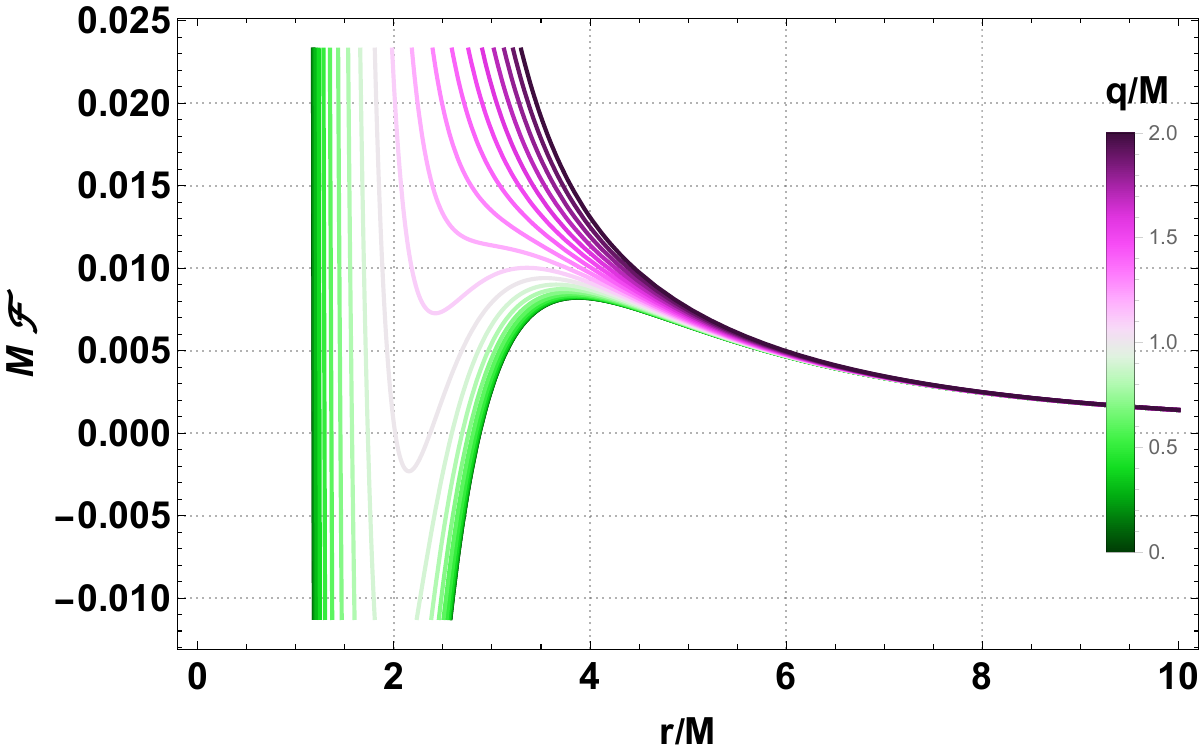}\\
     (i) $q=0.5$ \hspace{6cm} (ii) $\lambda=0.5$
    \caption{The dependence of the effective force $\mathcal{F}$ experienced by photon particles on radial coordinates $r$ for different values of $\lambda$ and $q$ for photons, while keeping $\gamma=1=M$ fixed.}
    \label{fig:force-null}
\end{figure*}

\begin{figure*}[tbhp]
    \centering
    \includegraphics[width=0.4\linewidth]{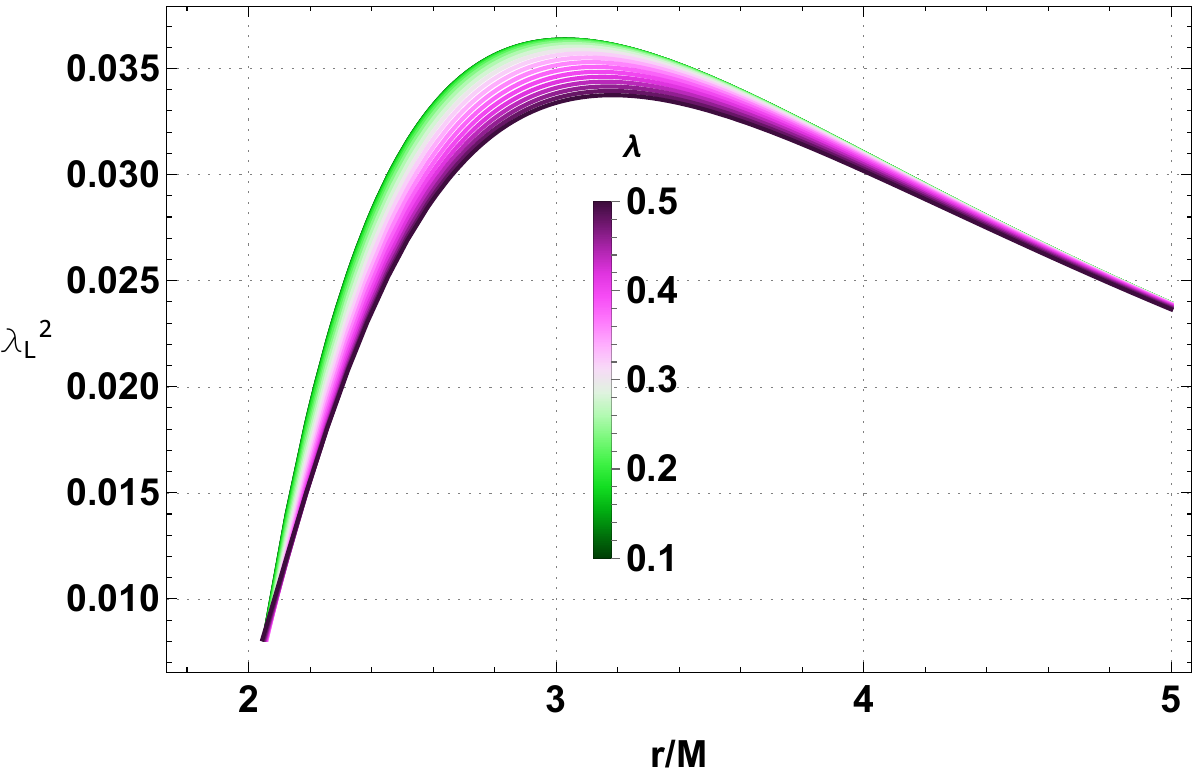}\qquad
    \includegraphics[width=0.4\linewidth]{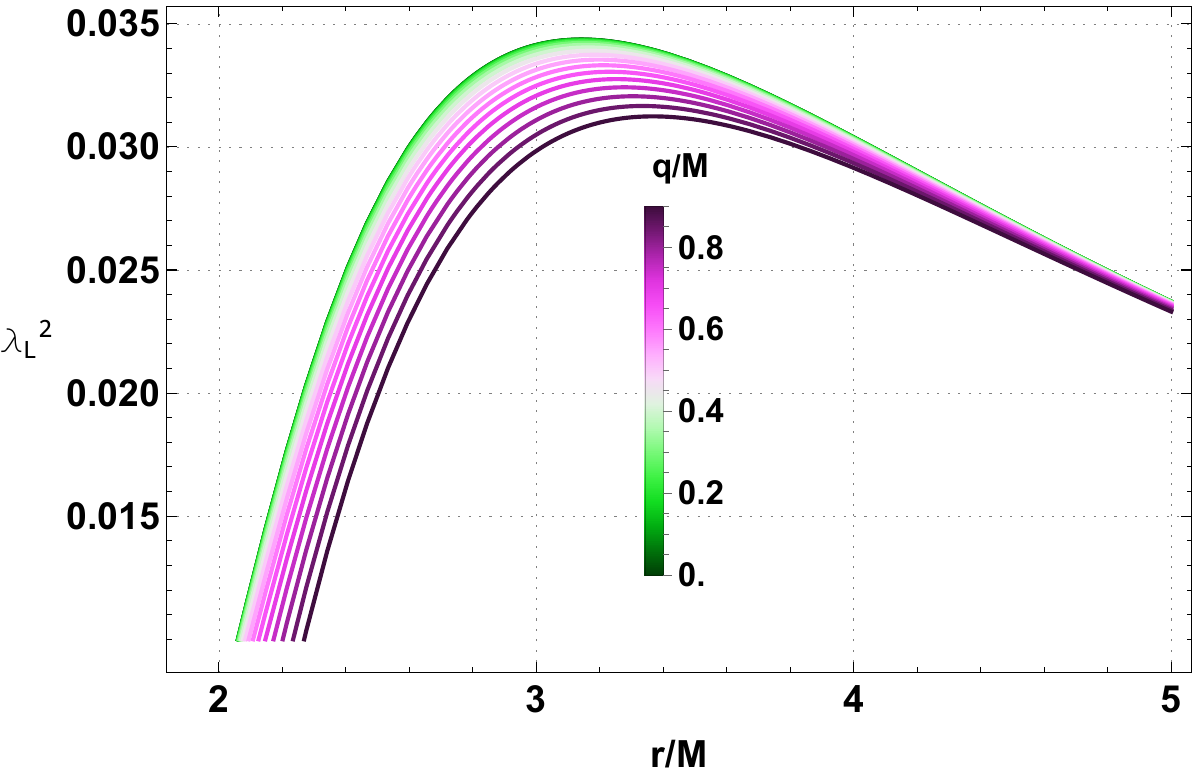}\\
    (i) $q/M=0.5$ \hspace{6cm} (ii) $\lambda=0.5$
    \caption{The dependence of the squared Lyapunov exponent $\lambda^2_{L}$ for various values of $\lambda$ and $q$. Here $\gamma=1=M$.}
    \label{fig:exponent}
\end{figure*}

\begin{figure*}[tbhp]
    \centering
    \includegraphics[width=0.3\linewidth]{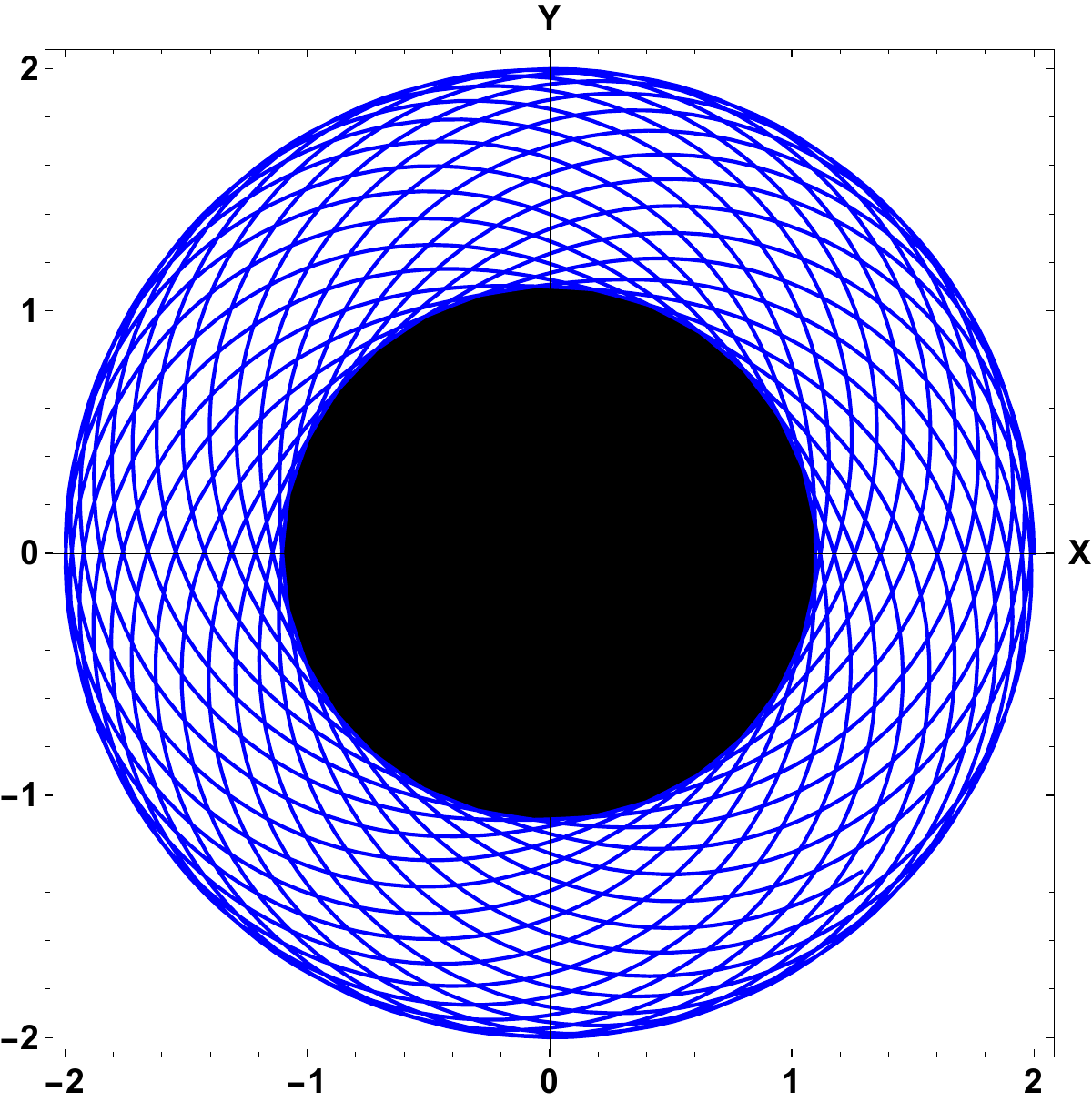}\quad
    \includegraphics[width=0.3\linewidth]{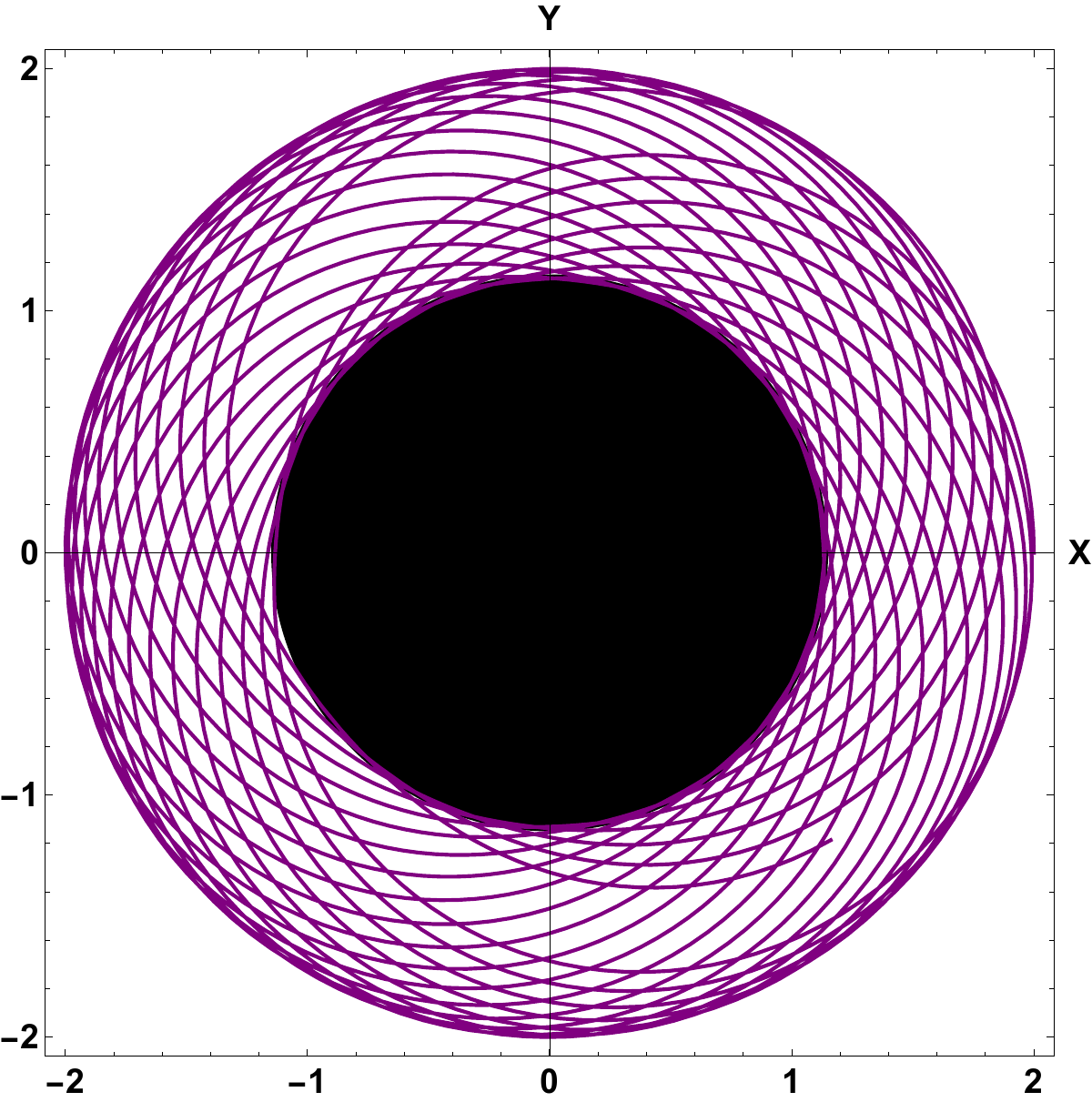}\quad
    \includegraphics[width=0.3\linewidth]{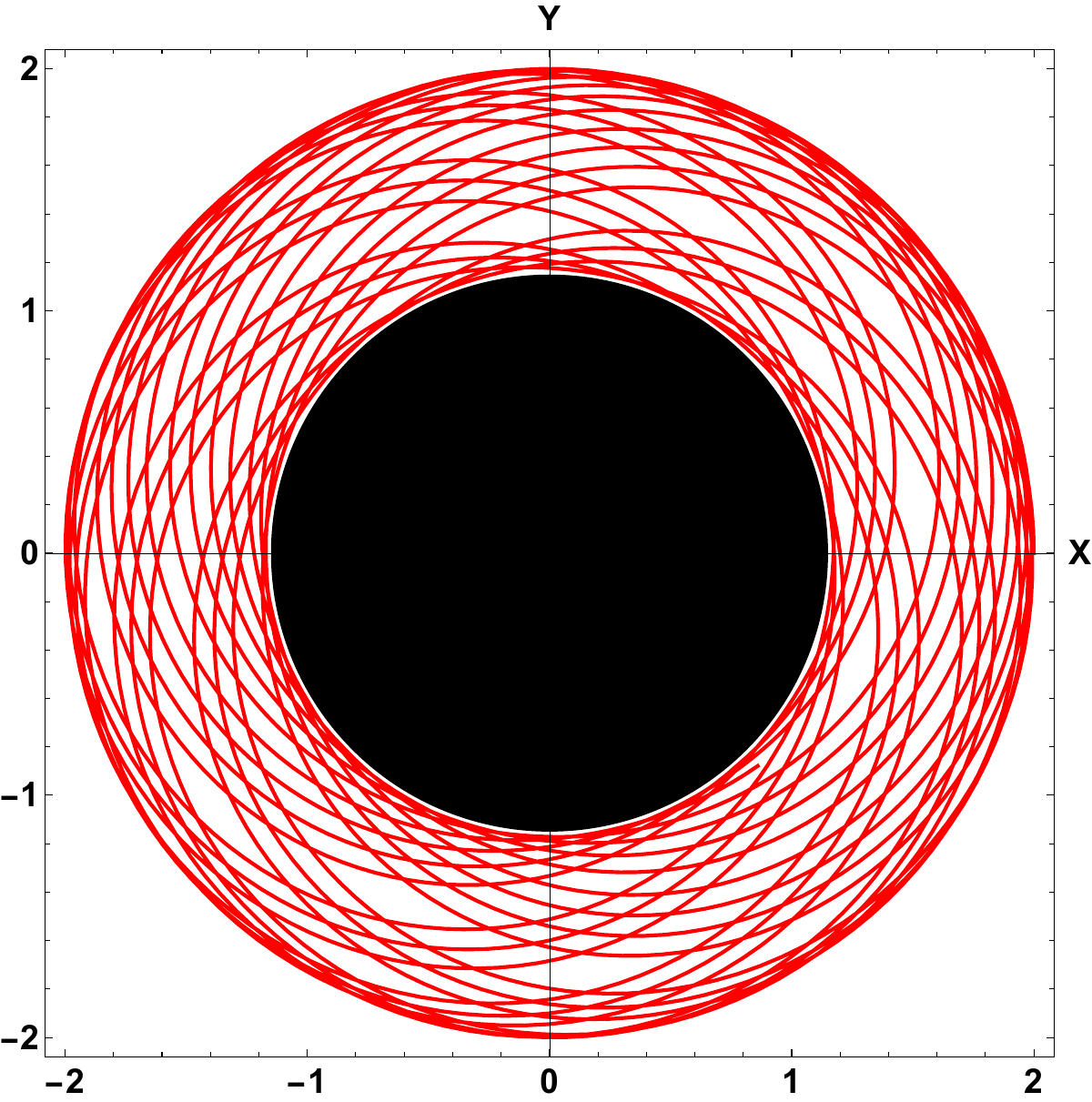}\\
    (i) $\lambda=0.5$ \hspace{4cm} (ii) $\lambda=0.6$ \hspace{4cm} (iii) $\lambda=0.7$
    \caption{Parametric plots of the photon trajectories, described by $u(\phi)=1/r(\phi)$, in the $X$--$Y$ plane for different values of the Kalb--Ramond field parameter $\lambda$, while keeping $\gamma = 1$, $q = 0.5$, and $M = 1$ fixed. The corresponding boundary conditions are chosen as $u(0)=0.5$ and $u'(0)=0$.}
    \label{fig:trajectory}
\end{figure*}

The radial profile of the force \(\mathcal{F}\) for different values of \(\lambda\) and \(q\) is displayed in Figure~\ref{fig:force-null}. Increasing either parameter enhances the effective force, indicating stronger gravitational influence in the near-horizon region. This corresponds to a deeper effective potential well, which increases the likelihood of photon capture and strengthens the trapping of null geodesics, thereby suppressing photon escape to infinity.

We now address the stability of circular null geodesics. Their instability timescale is set by the Lyapunov exponent~\cite{Cardoso2009}
\begin{equation}
\lambda^\text{null}_L=\sqrt{-\frac{V''_\text{eff}(r)}{2\dot{t}^2}}.\label{hh3}
\end{equation}

Using Eqs.~(\ref{gg3}), (\ref{gg5}), together with the circular orbit condition (\ref{photon1}) we obtain 
\begin{widetext}
\begin{align}
\lambda^\text{null}_L=\sqrt{f(r)\left(\frac{f(r)}{r^2}-\frac{f''(r)}{2}\right)}=\sqrt{1-\frac{2M r^2}{r^3+q^3}+\frac{\gamma}{r^{2/\lambda}}}
\sqrt{\frac{1}{r^2} - \frac{2M}{q^3 + r^3} + \frac{2M \left( q^6 - 7q^3 r^3 + r^6 \right)}{\left( q^3 + r^3 \right)^3}+ r^{-\frac{2(1+\lambda)}{\lambda}} \gamma - \frac{r^{-\frac{2(1+\lambda)}{\lambda}} \gamma (2 + \lambda)}{\lambda^2}}\Bigg|_{r=r_s}.\label{stable2}
\end{align}
\end{widetext}
Equation~(\ref{stable2}) demonstrates that the instability is governed by $q$, $\gamma$, $\lambda$, and $M$. In the limiting case $q=\gamma=0$, the standard Schwarzschild result is recovered.

Figure~\ref{fig:exponent} illustrates the behavior of the squared Lyapunov exponent, \(\lambda_L^2\), as a function of \(r\) for different values of \(\lambda\) and \(q\). Since \(\lambda_L^2>0\) throughout the plotted region, the corresponding circular null geodesics are unstable under radial perturbations. The curves attain a maximum at intermediate \(r\), indicating the region of strongest instability, and decrease at larger distances, implying weaker instability far from the black hole. Moreover, increasing \(\lambda\) and \(q\) lowers \(\lambda_L^2\), corresponding to a longer instability timescale and comparatively less unstable null orbits.

\subsection{Eikonal QNMs}\label{isec4-QNM}

The instability rate $\lambda^{\rm null}_L$ and the angular frequency at the photon sphere
\begin{equation}
\Omega_c=\left.\frac{\sqrt{f(r)}}{r}\right|_{r=r_s}\label{Omegac}
\end{equation}
fix the eikonal QNM frequencies for high-multipole scalar perturbations \cite{Cardoso2009},
\begin{equation}
\omega_{\rm eik}=(\ell+1/2)\,\Omega_c-i(n+1/2)\,|\lambda^{\rm null}_L|+\mathcal{O}(\ell^{-1}).\label{eikonal}
\end{equation}
This correspondence has known caveats outside General Relativity \cite{WOS:000406183300090}; for the present metric it remains controlled because the geometry is asymptotically flat in the EHT-allowed window. Sample values at $\ell=2$, $n=0$, $\lambda=0.5$, $M=1$ read $(\Omega_c M,|\lambda_L|M)\simeq(0.193,0.271)$ for $(q,\gamma)=(0,0.1)$ and $(0.199,0.249)$ for $(q,\gamma)=(0.8,0.5)$, giving ${\rm Re}(\omega)M\simeq 0.48$--$0.50$ and $|{\rm Im}(\omega)|M\simeq 0.13$--$0.12$ across the table. The real part rises and the damping drops as $q$ and $\gamma$ grow, a feature future ground-based gravitational-wave observatories could test \cite{WOS:000279179100001}.

\subsection{Orbit equation and photon trajectories}

To further investigate photon dynamics, we derive the orbit equation by combining Eqs.~(\ref{gg3})-- (\ref{gg5}) yielding
\begin{equation}
\frac{\dot{r}^2}{\dot{\phi}^2}=\left(\frac{dr}{d\phi}\right)^2=r^4\left[\frac{1}{\beta^2}-\frac{1}{r^2}\left(1-\frac{2M r^2}{r^3+q^3}+\frac{\gamma}{r^{2/\lambda}}\right)\right],\label{hh4}
\end{equation}
where  $\beta=\frac{\mathrm{L}}{\mathrm{E}}$ denotes the impact parameter associated with photon trajectories.

Introducing the standard transformation $u=\frac{1}{r}$, we obtain the trajectory equation:
\begin{equation}
   \left(\frac{du}{d\phi}\right)^2+u^2 = \frac{1}{\beta_c^2}  + \frac{2M u^3}{1 + q^3 u^3} - \gamma \, u^{2\left(1+\frac{1}{\lambda}\right)}.\label{hh5}
\end{equation}
Equation~(\ref{hh5}) describes the propagation of photons in the gravitational background of the KR black hole coupled to nonlinear electrodynamics. The resulting photon trajectories are significantly influenced by the magnetic charge $q$, the Lorentz-violating parameters $\gamma$ and $\lambda$, and the black-hole mass $M$. These quantities collectively determine the structure of null geodesics and the corresponding optical properties of the spacetime.

Differentiating both sides of Eq.~(\ref{hh5}) with respect to $\phi$ and after re-arranging, we find the following second-order differential equation
\begin{equation}
\frac{d^2u}{d\phi^2}+u=\frac{3Mu^2}{(1+q^3u^3)^2}-\gamma\left(1+\frac{1}{\lambda}\right)u^{1+\frac{2}{\lambda}} .\label{hh6b}
\end{equation}

In Figure~\ref{fig:trajectory}, we illustrate the photon trajectories, described by $u(\phi)=1/r(\phi)$, in the $X$--$Y$ plane for different values of the KR field parameter $\lambda$, while keeping $\gamma = 1$, $q = 0.5$, and $M = 1$ fixed. The corresponding boundary conditions are taken as $u(0)=0.5$ and $u'(0)=0$. From the figure, it is observed that the photon trajectory pattern changes noticeably with increasing values of $\lambda$, indicating that the KR field parameter has a significant influence on the propagation and bending behavior of photons around the BH spacetime. Similarly, one can depict the photon trajectories by varying the electric charge $q$ for particular values of $\lambda$.

\section{Hawking temperature and sparsity}\label{sec5}

A static, spherically symmetric horizon at $r=r_h$ defined by $f(r_h)=0$ carries Hawking temperature
\begin{equation}
T_H=\frac{f'(r_h)}{4\pi}=\frac{1}{4\pi}\!\left[\frac{2M(r_h^4-2r_h q^3)}{(r_h^3+q^3)^2}-\frac{2\gamma}{\lambda\, r_h^{2/\lambda+1}}\right],\label{TH}
\end{equation}
and horizon area $A_H=4\pi r_h^2$. We compare these against the Schwarzschild values $T_H^{\rm Sch}=1/(8\pi M)$ and $A_H^{\rm Sch}=16\pi M^2$.

The Hawking cascade from astrophysical black holes is famously sparse: consecutive Hawking quanta are emitted on timescales much longer than the characteristic period of the quanta themselves \cite{WOS:000377442000006}. The Gray--Visser sparsity parameter,
\begin{equation}
\eta_{\rm sp}=\frac{\lambda_{\rm th}^2}{A_H}=\frac{(2\pi/T_H)^2}{4\pi r_h^2}=\frac{\pi}{T_H^2\,r_h^2},\label{eta}
\end{equation}
captures how many thermal wavelengths $\lambda_{\rm th}=2\pi/T_H$ fit per unit horizon area. Schwarzschild gives $\eta_{\rm sp}^{\rm Sch}=16\pi^3\simeq 496$, the canonical benchmark \cite{WOS:000377442000006}. A value of $\eta_{\rm sp}\gg 1$ signals an emission process where photons leave one at a time; the larger $\eta_{\rm sp}$, the sparser the cascade.

Table~\ref{tab:hawking} reports $r_h$, $T_H$, $A_H$ and $\eta_{\rm sp}$ at $\lambda=0.5$, $M=1$. Both $T_H$ and $A_H$ fall as $q$ or $\gamma$ grow, while $\eta_{\rm sp}$ climbs. The combined effect of magnetic charge and KR-induced Lorentz breaking pushes $\eta_{\rm sp}$ from the Schwarzschild value up to roughly $1.7\times 10^{3}$ at $(q,\gamma)=(0.8,0.5)$. The cascade is then about $3.5\times$ sparser than for Schwarzschild.

\begin{table}[ht!]
\centering
\renewcommand{\arraystretch}{1.45}
\setlength{\tabcolsep}{6pt}
\begin{tabular}{|c|c|c|c|c|c|}
\hline
$q$ & $\gamma$ & $r_h$ & $T_H\,M$ & $A_H/M^2$ & $\eta_{\rm sp}$ \\ \hline
0.0 & 0.1 & 1.9873 & 0.03927 & 49.63 & 515.8 \\ \hline
0.0 & 0.3 & 1.9602 & 0.03812 & 48.28 & 562.6 \\ \hline
0.0 & 0.5 & 1.9305 & 0.03677 & 46.83 & 623.5 \\ \hline
0.4 & 0.1 & 1.9703 & 0.03857 & 48.79 & 543.9 \\ \hline
0.4 & 0.3 & 1.9417 & 0.03730 & 47.38 & 598.9 \\ \hline
0.4 & 0.5 & 1.9100 & 0.03579 & 45.85 & 672.4 \\ \hline
0.8 & 0.1 & 1.8293 & 0.03217 & 42.05 & 907.0 \\ \hline
0.8 & 0.3 & 1.7805 & 0.02921 & 39.84 & 1161.6 \\ \hline
0.8 & 0.5 & 1.7179 & 0.02487 & 37.09 & 1721.4 \\ \hline
\end{tabular}
\caption{Horizon radius $r_h$, Hawking temperature $T_H$, horizon area $A_H$ and Gray--Visser sparsity parameter $\eta_{\rm sp}$, at $\lambda=0.5$, $M=1$. Schwarzschild reference: $T_H^{\rm Sch}=1/(8\pi)\simeq 0.0398$, $A_H^{\rm Sch}=16\pi\simeq 50.27$, $\eta_{\rm sp}^{\rm Sch}=16\pi^3\simeq 496$.}
\label{tab:hawking}
\end{table}

The physical reading is direct. The KR sector and the magnetic charge cool the horizon and lower its area; the thermal wavelength $\lambda_{\rm th}=2\pi/T_H$ grows accordingly. Because $\eta_{\rm sp}\propto T_H^{-2}r_h^{-2}$, both effects compound: $\eta_{\rm sp}$ scales as the inverse fourth power of typical horizon quantities. The cascade therefore becomes sparser at a faster rate than $T_H$ alone falls. We expected the dependence on $\gamma$ to be approximately linear at small $\gamma$, but it is not; the deviation is already a few percent at $\gamma=0.3$.
 
\section{Conclusion}\label{sec6}

We have studied the geodesic structure and the optical and thermal properties of a static, spherically symmetric black hole sourced by a Kalb--Ramond field coupled to nonlinear electrodynamics, parameterized by the mass $M$, the magnetic monopole charge $q$ and the Lorentz-violating parameters $(\gamma,\lambda)$. The metric (\ref{function}) reduces to Schwarzschild in the limit $q,\gamma\to 0$, to the Hayward solution at $\gamma,\lambda\to 0$, and to a Reissner--Nordstr\"om-like form (with $\gamma$ acting as Lorentz-violating hair rather than electric charge) at $q=0$, $\lambda=1$.

Several specific results follow. First, the ISCO retreats from the Schwarzschild value $6M$ as either $q$ or $\gamma$ grows, the effect being stronger in $q$; the orbital frequency $\Omega_\phi$ grows with $q$ but falls with $\lambda$, the two acting as opposite knobs on the rotational motion. Second, the photon sphere and the shadow shrink with $q$ and $\gamma$, with the Schwarzschild values $r_s=3M$ and $R_{\rm sh}=3\sqrt{3}\,M$ correctly recovered. Third, every $(q,\gamma)$ entry in Table~\ref{tab:photon-radius} satisfies both the M87$^\ast$ and Sgr~A$^\ast$ 1$\sigma$ EHT bands; the Sgr~A$^\ast$ data is the more restrictive of the two and starts to bound $\lambda$ from above once $\gamma\gtrsim 0.3$ (Table~\ref{tab:EHT-window}). Fourth, the Lyapunov exponent for unstable null circular orbits feeds directly into the eikonal QNM spectrum $\omega_{\rm eik}=(\ell+1/2)\Omega_c-i(n+1/2)|\lambda_L|$, and we have given representative numerical values. Fifth, the Hawking temperature and area both decrease with $q$ and $\gamma$, while the Gray--Visser sparsity parameter $\eta_{\rm sp}$ climbs from the Schwarzschild value $16\pi^3\simeq 496$ to $\simeq 1721$ at $(q,\gamma)=(0.8,0.5)$; the Hawking cascade is therefore about $3.5\times$ sparser than for Schwarzschild, and consecutive emissions are even more widely separated in time.

Taken together, the results identify a black hole geometry that remains observationally viable in the strong-field regime probed by the EHT, modifies the eikonal QNM spectrum by an amount detectable by future gravitational-wave observatories such as LISA and the Einstein Telescope, and produces a Hawking emission noticeably less continuous than that of Schwarzschild. A natural follow-up is the slowly rotating extension obtained through a Newman--Janis-type construction; another is the computation of the greybody factors and the time-domain QNM profile beyond the eikonal limit, with a view to comparing against future LISA-band signals.

\section*{Acknowledgments}  

F.~A.\ acknowledges the Inter University Centre for Astronomy and Astrophysics (IUCAA), Pune, India, for granting a visiting associateship. \.{I}.~S.\ expresses gratitude to T\"{U}B\.{I}TAK, ANKOS, and SCOAP3 for helpful support. He also acknowledges COST Actions CA22113, CA21106, CA23130, CA21136, and CA23115 for their contributions to networking.

\section*{ Data availability statement}

There is no data in this manuscript. Computational verification scripts are available from the authors on request.

\section*{Conflict of Interests}

The authors declare no conflict of interests.

\bibliographystyle{apsrev4-1}

\bibliography{references}

@article{WOS:000377442000006,
    author = {Gray, Finnian and Schuster, Sebastian and Van-Brunt, Alexander and Visser, Matt},
    title = {The Hawking cascade from a black hole is extremely sparse},
    journal = {Class. Quantum Grav.},
    year = {2016},
    volume = {33},
    pages  ={115003},
    number = {11},
    doi = {10.1088/0264-9381/33/11/115003},
    url = {https://doi.org/10.1088/0264-9381/33/11/115003},
}

@article{ WOS:000279179100001,
    author = {Stefanov, Ivan Zh. and Yazadjiev, Stoytcho S. and Gyulchev, Galin G.},
    title = {Connection between Black-Hole Quasinormal Modes and Lensing in the Strong Deflection Limit},
    journal = {Phys. Rev. Lett.},
    year = {2010},
    volume = {104},
    pages  = {251103},
    number = {25},
    doi = {10.1103/PhysRevLett.104.251103},
    url = {https://doi.org/10.1103/PhysRevLett.104.251103},
}

@article{ WOS:000406183300090,
    author = {Konoplya, R. A. and Stuchlik, Z.},
    title = {Are eikonal quasinormal modes linked to the unstable circular null geodesics?},
    journal = {Phys. Lett. B},
    year = {2017},
    volume = {771},
    pages = {597-602},
    doi = {10.1016/j.physletb.2017.06.015},
    url = {https://doi.org/10.1016/j.physletb.2017.06.015},
}

@article{ WOS:000571990600005,
    author = {Kocherlakota, Prashant and Rezzolla, Luciano},
    title = {Accurate mapping of spherically symmetric black holes in a parametrized framework},
    journal = {Phys. Rev. D},
    year = {2020},
    volume = {102},
    pages  ={064058},
    number = {6},
    doi = {10.1103/PhysRevD.102.064058},
    url = {https://doi.org/10.1103/PhysRevD.102.064058},
}

@article{Singh2026,
    author = {D. V. Singh and S. Upadhyay and P. Paul and K. Myrzakulov},
    title = {Extended thermodynamics and $P-v$ Criticality of Kalb-Ramond black hole coupled with nonlinear electrodynamics},
    journal = {Eur. Phys. J Plus},
    volume = {141},
    pages = {364},
    year = {2026},
    doi   ={https://doi.org/10.1140/epjp/s13360-026-07612-w}
}

@article{BornInfeld1934,
  author  = {Born, Max and Infeld, Leopold},
  title   = {{Foundations of the new field theory}},
  journal = {Proc. Roy. Soc. Lond. A},
  volume  = {144},
  pages   = {425},
  year    = {1934},
  doi     = {10.1098/rspa.1934.0059}
}

@article{Altschul2010,
  author  = {B. Altschul and Q. G. Bailey and V. A. Kosteleck{\'y}},
  title   = {Lorentz violation with an antisymmetric tensor},
  journal = {Phys. Rev. D},
  volume  = {81},
  pages   = {065028},
  year    = {2010},
  doi     = {https://doi.org/10.1103/PhysRevD.81.065028}
}

@book{Green1987,
  author    = {Michael B. Green and John H. Schwarz and Edward Witten},
  title     = {Superstring Theory, Volume 2},
  year      = {1987},
  publisher = {Cambridge University Press},
  address   = {Cambridge}
}

@article{Hehl1976,
  author  = {F. W. Hehl and P. von der Heyde and G. D. Kerlick and J. M. Nester},
  title   = {General relativity with spin and torsion: Foundations and prospects},
  journal = {Rev. Mod. Phys.},
  volume  = {48},
  pages   = {393},
  year    = {1976},
  doi     = {https://doi.org/10.1103/RevModPhys.48.393}
}

@article{Hehl1995,
  author  = {F. W. Hehl and P. von der Heyde and G. D. Kerlick and J. M. Nester},
  title   = {Metric affine gauge theory of gravity: Field equations, Noether identities, world spinors, and breaking of dilation invariance},
  journal = {Phys. Rep.},
  volume  = {258},
  pages   = {1},
  year    = {1995},
  doi     = {https://doi.org/10.1016/0370-1573(94)00111-F}
}

@article{Ayon-Beato:2000mjt,
  author  = {Eloy Ayon-Beato and Alberto Garcia},
  title   = {The Bardeen model as a nonlinear magnetic monopole},
  journal = {Phys. Lett, B},
  volume  = {493},
  pages   = {149--152},
  year    = {2000},
  doi     = {https://doi.org/10.1016/S0370-2693(99)01038-2}
}

@article{Ayon-Beato:1999qin,
  author  = {Eloy Ayon-Beato and Alberto Garcia},
  title   = {Nonsingular charged black hole solution for nonlinear source},
  journal = {Gen. Relativ. Gravit.},
  volume  = {31},
  pages   = {629--633},
  year    = {1999},
  doi     = {https://doi.org/10.1023/A:1026640911319}
}

@article{Ayon-Beato:1998hmi,
  author  = {Eloy Ayon-Beato and Alberto Garcia},
  title   = {Regular black hole in general relativity coupled to nonlinear electrodynamics},
  journal = {Phys. Rev. Lett.},
  volume  = {80},
  pages   = {5056--5059},
  year    = {1998},
  doi     = {https://doi.org/10.1103/PhysRevLett.80.5056}
}

@article{Reissner1916,
  author  = {H. Reissner},
  title   = {{\"U}ber die Eigengravitation des elektrischen Feldes nach der Einsteinschen Theorie},
  journal = {Annalen der Physik},
  volume  = {50},
  pages   = {106--120},
  year    = {1916},
  doi     = {https://doi.org/10.1002/andp.19163550905}
  
}

@article{Nordstrom1918,
  author  = {G. Nordstr{\"o}m},
  title   = {On the Energy of the Gravitation field in Einstein's Theory},
  journal = {Proceedings of the Royal Netherlands Academy of Arts and Sciences},
  volume  = {20},
  pages   = {1238--1245},
  year    = {1918}
}

@article{Nandi2023,
  author  = {K. K. Nandi and R. N. Izmailov and R. Kh. Karimov and A. A. Potapov},
  title   = {On the Kalb--Ramond modified Lorentz violating hairy black holes and Thorne's hoop conjecture},
  journal = {Eur. Phys. J C},
  volume  = {83},
  pages   = {984},
  year    = {2023},
  doi     = {https://doi.org/10.1140/epjc/s10052-023-12172-9}
}

@article{Penrose1965,
    author = {Roger Penrose},
    title = {Gravitational Collapse and Space-Time Singularities},
    journal = {Phys. Rev. Lett.},
    volume = {14},
    number = {3},
    pages = {57--59},
    year = {1965},
    doi = {10.1103/PhysRevLett.14.57},
    url = {https://doi.org/10.1103/PhysRevLett.14.57}
}

@article{Hawking1970,
    author = {S. W. Hawking and R. Penrose},
    title = {The Singularities of Gravitational Collapse and Cosmology},
    journal = {Proc. R. Soc. London, Series A: Math. Phys. Sci.},
    volume = {314},
    number = {1519},
    pages = {529--548},
    year = {1970},
    doi = {10.1098/rspa.1970.0021},
    url = {https://doi.org/10.1098/rspa.1970.0021}
}

@article{Synge1966,
    author = {J. L. Synge},
    title = {The Escape of Photons from Gravitationally Intense Stars},
    journal = {MNRAS},
    volume = {131},
    number = {3},
    pages = {463--466},
    year = {1966},
    doi = {10.1093/mnras/131.3.463},
    url = {https://doi.org/10.1093/mnras/131.3.463}
}

@article{Bardeen1972,
    author = {J. M. Bardeen and W. H. Press and S. A. Teukolsky},
    title = {Rotating Black Holes: Locally Nonrotating Frames, Energy Extraction, and Scalar Synchrotron Radiation},
    journal = {Astrophys. J},
    volume = {178},
    pages = {347--370},
    year = {1972},
    doi = {10.1086/151796},
    url = {https://doi.org/10.1086/151796}
}

@article{Luminet1979,
    author = {Jean-Pierre Luminet},
    title = {Image of a Spherical Black Hole with Thin Accretion Disk},
    journal = {Astron. Astrophys.},
    volume = {75},
    pages = {228--235},
    year = {1979},
    url = {https://articles.adsabs.harvard.edu/pdf/1979A%26A....75..228L}
}

@article{Cardoso2009,
  author  = {Vitor Cardoso and Alex S. Miranda and Emanuele Berti and Helvi Witek and Vilson T. Zanchin},
  title   = {Geodesic Stability, Lyapunov Exponents, and Quasinormal Modes},
  journal = {Phys. Rev. D},
  volume  = {79},
  number  = {6},
  pages   = {064016},
  year    = {2009},
  doi     = {10.1103/PhysRevD.79.064016},
  url     = {https://doi.org/10.1103/PhysRevD.79.064016}
}

@article{Lessa2021,
    author = {L. A. Lessa and R. Oliveira and J. E. G. Silva and C. A. S. Almeida},
    title = {Traversable Wormhole Solution with a Background Kalb--Ramond Field},
    journal = {Annals of Physics},
    volume = {433},
    pages = {168604},
    year = {2021},
    month = oct,
    doi = {10.1016/j.aop.2021.168604},
    url = {https://doi.org/10.1016/j.aop.2021.168604}
}

@article{Jumaniyozov2025,
    author = {Jumaniyozov, S. and Murodov, S. and Rayimbaev, J. and others},
    title = {Black holes surrounded by PFDM in Kalb-Ramond gravity: from thermodynamics to QPO tests},
    journal = {Eur. Phys. J C},
    volume = {85},
    pages = {797},
    year = {2025},
    doi = {10.1140/epjc/s10052-025-14522-1},
    url = {https://doi.org/10.1140/epjc/s10052-025-14522-1}
}

@article{Gullu2022,
    author = {G{\"u}ll{\"u}, {\.I}brahim and {\"O}vg{\"u}n, Ali},
    title = {Schwarzschild-like black hole with a topological defect in bumblebee gravity},
    journal = {Ann. Phys. (NY)},
    volume = {436},
    pages = {168721},
    year = {2022}
}

@article{Fathi2025,
    author = {Fathi, M. and {\"O}vg{\"u}n, A.},
    title = {Black hole with global monopole charge in self-interacting Kalb-Ramond field},
    journal = {Eur. Phys. J Plus},
    volume = {140},
    pages = {280},
    year = {2025},
    doi = {10.1140/epjp/s13360-025-06241-z},
    url = {https://doi.org/10.1140/epjp/s13360-025-06241-z}
}

@misc{Faizuddin2026b,
    author = {Ahmed, Faizuddin and Fathi, Mohsen and Silva, Edilberto O.},
    year = {2026},
    eprint = {2604.11357},
    archivePrefix= {arXiv},
    primaryClass = {gr-qc},
    doi = {10.48550/arXiv.2604.11357},
    url = {https://doi.org/10.48550/arXiv.2604.11357}
}

@misc{Silva2025,
    author = {Ahmed, F. and Silva, E. O.},
    year = {2025},
    eprint = {2511.21374},
    archivePrefix= {arXiv},
    primaryClass = {hep-th},
    doi={https://doi.org/10.48550/arXiv.2511.21374}
}

@article{Baruah2025,
    author = {Baruah, A. and Sekhmani, Y. and Maurya, S. K. and Deshamukhya, A. and Jasim, M. K.},
    title = {Quasinormal modes, greybody factors, and Hawking radiation sparsity of black holes influenced by a global monopole charge in Kalb-Ramond gravity},
    journal = {JCAP},
    volume = {2025},
    number = {08},
    pages = {023},
    year = {2025},
    doi = {10.1088/1475-7516/2025/08/023},
    url = {https://doi.org/10.1088/1475-7516/2025/08/023}
}

@misc{Ahmad2026,
    author = {Ahmed, Faizuddin and Al-Badawi, Ahmad and Sakall{\i}, {\.I}zzet},
    year = {2026},
    eprint = {2601.10303},
    archivePrefix= {arXiv},
    primaryClass = {gr-qc},
    doi = {10.48550/arXiv.2601.10303},
    url = {https://doi.org/10.48550/arXiv.2601.10303}
}

@misc{Badawi2026,
    author = {Ahmed, Faizuddin and Al-Badawi, Ahmad and Silva, Edilberto O.},
    year = {2026},
    eprint = {2602.15570},
    archivePrefix= {arXiv},
    primaryClass = {gr-qc},
    doi = {10.48550/arXiv.2602.15570},
    url = {https://doi.org/10.48550/arXiv.2602.15570}
}

@misc{AhmedSilva2026,
    author = {Ahmed, Faizuddin and Silva, Edilberto O.},
    year = {2026},
    eprint = {2603.11312},
    archivePrefix= {arXiv},
    primaryClass = {gr-qc},
    doi = {10.48550/arXiv.2603.11312},
    url = {https://doi.org/10.48550/arXiv.2603.11312}
}

@article{Atamurotov2022,
  author    = {Atamurotov, F. and Ortiqboev, D. and Abdujabbarov, A. and others},
  title     = {Particle dynamics and gravitational weak lensing around black hole in the Kalb-Ramond gravity},
  journal   = {Eur. Phys. J C},
  volume    = {82},
  pages    = {659},
  year      = {2022},
  doi       = {10.1140/epjc/s10052-022-10619-z},
  url       = {https://doi.org/10.1140/epjc/s10052-022-10619-z}
}

@article{Dymnikova2002,
    author = {Dymnikova, I.},
    title = {Regular electrically charged structures in nonlinear electrodynamics coupled to general relativity},
    journal = {Class. Quantum Grav.},
    volume = {19},
    pages = {725--740},
    year = {2002},
    doi = {10.1088/0264-9381/19/4/310},
    url = {https://doi.org/10.1088/0264-9381/19/4/310}
}

@article{Dymnikova2004,
    author = {Dymnikova, I.},
    title = {Regular rotating electrically charged black holes and solitons in nonlinear electrodynamics minimally coupled to gravity},
    journal = {Class. Quantum Grav.},
    volume = {21},
    pages = {4417--4429},
    year = {2004},
    doi = {10.1088/0264-9381/21/18/009},
    url = {https://doi.org/10.1088/0264-9381/21/18/009}
}

@article{FanWang2016,
  author  = {Fan, Z. Y. and Wang, Xiaobao},
  title   = {Construction of Regular Black Holes in General Relativity},
  journal = {Phys. Rev. D},
  volume  = {94},
  number  = {12},
  pages   = {124027},
  year    = {2016},
  doi     = {10.1103/PhysRevD.94.124027},
  url     = {https://doi.org/10.1103/PhysRevD.94.124027}
}

@article{Rodrigues2016,
    author = {Rodrigues, M. E. and Junior, E. L. B. and Marques, G. T. and Zanchin, V. T.},
    title = {Regular black holes in $f(R)$ gravity coupled to nonlinear electrodynamics},
    journal = {Phys. Rev. D},
    volume = {94},
    number = {2},
    pages = {024062},
    year = {2016},
    doi = {10.1103/PhysRevD.94.024062},
    url = {https://doi.org/10.1103/PhysRevD.94.024062}
}

@article{Balart2014,
    author = {Balart, Leonardo and Vagenas, Elias C.},
    title = {Regular black holes with a nonlinear electrodynamics source},
    journal = {Phys. Rev. D},
    volume = {90},
    pages = {124045},
    year = {2014},
    doi = {10.1103/PhysRevD.90.124045},
    url = {https://doi.org/10.1103/PhysRevD.90.124045}
}

@article{Atamurotov2022KRBH,
  author  = {Atamurotov, Farruh and Ortiqboev, Dilshod and Abdujabbarov, Ahmadjon and others},
  title   = {Particle dynamics and gravitational weak lensing around black hole in the Kalb--Ramond gravity},
  journal = {Eur. Phys. J C},
  volume  = {82},
  pages   = {659},
  year    = {2022},
  doi     = {10.1140/epjc/s10052-022-10619-z},
  url     = {https://doi.org/10.1140/epjc/s10052-022-10619-z}
}

@article{Ma2015,
  author  = {Ma, M.-S.},
  title   = {Magnetically charged regular black hole in a model of nonlinear electrodynamics},
  journal = {Ann. Phys. (NY)},
  volume  = {362},
  pages   = {529--537},
  year    = {2015},
  doi     = {10.1016/j.aop.2015.08.028},
  url     = {https://doi.org/10.1016/j.aop.2015.08.028}
}

@book{Chandrasekhar1983,
  author    = {Chandrasekhar, S.},
  title     = {The Mathematical Theory of Black Holes},
  year      = {1983},
  publisher = {Oxford University Press},
  address   = {Oxford}
}

@article{He2022,
    author = {He, A. and Tao, J. and Wang, P. and Xue, Y. and Zhang, L.},
    title = {Effects of Born--Infeld electrodynamics on black hole shadows},
    journal = {Eur. Phys. J C},
    volume = {82},
    pages = {683},
    year = {2022},
    doi = {10.1140/epjc/s10052-022-10637-x},
    url = {https://doi.org/10.1140/epjc/s10052-022-10637-x}
}

@article{Allahyari2020,
    author = {Allahyari, A. and Khodadi, M. and Vagnozzi, S. and Mota, D. F.},
    title = {Magnetically charged black holes from non-linear electrodynamics and the Event Horizon Telescope},
    journal = {JCAP},
    volume = {2020},
    number = {02},
    pages = {003},
    year = {2020},
    doi = {10.1088/1475-7516/2020/02/003},
    url = {https://doi.org/10.1088/1475-7516/2020/02/003}
}

@article{Kruglov2015,
    author = {Kruglov, S. I.},
    title = {A model of nonlinear electrodynamics},
    journal = {Ann. Phys. (NY)},
    volume = {353},
    pages = {299--306},
    year = {2015},
    doi = {10.1016/j.aop.2014.12.001},
    url = {https://doi.org/10.1016/j.aop.2014.12.001}
}

@article{Kruglov2016,
    author = {Kruglov, S. I.},
    title = {Modified Nonlinear Model of Arcsin-Electrodynamics},
    journal = {Commun. Theor. Phys.},
    volume = {66},
    number = {1},
    pages = {59},
    year = {2016},
    doi = {10.1088/0253-6102/66/1/059},
    url = {https://doi.org/10.1088/0253-6102/66/1/059}
}

@article{Kruglov2015b,
    author = {Kruglov, Sergey I.},
    title = {Nonlinear arcsin-electrodynamics},
    journal = {Ann. Physik (Berlin)},
    volume = {527},
    number = {5--6},
    pages = {397--401},
    year = {2015},
    doi = {10.1002/andp.201500142},
    url = {https://doi.org/10.1002/andp.201500142}
}

@article{Kruglov2017,
    author = {Kruglov, S. I.},
    title = {Magnetized black holes and nonlinear electrodynamics},
    journal = {Int. J Mod. Phys. A},
    volume = {32},
    number = {23--24},
    pages = {1750147},
    year = {2017},
    doi = {10.1142/S0217751X17501470},
    url = {https://doi.org/10.1142/S0217751X17501470}
}

@article{Kruglov2018,
    author = {Kruglov, S. I.},
    title = {Magnetically charged black hole in framework of nonlinear electrodynamics model},
    journal = {Int. J Mod. Phys. A},
    volume = {33},
    pages = {1850023},
    year = {2018},
    doi = {10.1142/S0217751X18500239},
    url = {https://doi.org/10.1142/S0217751X18500239}
}

@article{Kruglov2019,
    author = {Kruglov, S. I.},
    title = {Dyonic Black Holes with Nonlinear Logarithmic Electrodynamics},
    journal = {Gravit. Cosmol.},
    volume = {25},
    pages = {190--195},
    year = {2019},
    doi = {10.1134/S0202289319020105},
    url = {https://doi.org/10.1134/S0202289319020105}
}

@article{Kruglov2020,
    author = {Kruglov, S. I.},
    title = {Dyonic and magnetized black holes based on nonlinear electrodynamics},
    journal = {Eur. Phys. J C},
    volume = {80},
    pages = {250},
    year = {2020},
    doi = {10.1140/epjc/s10052-020-7809-x},
    url = {https://doi.org/10.1140/epjc/s10052-020-7809-x}
}

@article{Dymnikova2021,
    author = {Dymnikova, I. and Galaktionov, E.},
    title = {Regular electrically charged objects in nonlinear electrodynamics coupled to gravity},
    journal = {J. Phys.: Conf. Series},
    volume = {2103},
    pages = {012078},
    year = {2021},
    doi = {10.1088/1742-6596/2103/1/012078},
    url = {https://doi.org/10.1088/1742-6596/2103/1/012078}
}

@article{BornInfeld1933,
    author = {Born, M. and Infeld, L.},
    title = {Foundations of the new field theory},
    journal = {Nature},
    volume = {132},
    pages = {1004},
    year = {1933},
    doi = {10.1038/1321004a0}
}

@article{Bronnikov2001,
  author  = {Bronnikov, K. A.},
  title   = {Regular Magnetic Black Holes and Monopoles from Nonlinear Electrodynamics},
  journal = {Phys. Rev. D},
  volume  = {63},
  pages   = {044005},
  year    = {2001},
  doi     = {10.1103/PhysRevD.63.044005},
  url     = {https://doi.org/10.1103/PhysRevD.63.044005}
}

@article{Gullu2021,
  author  = {G{\"u}ll{\"u}, I. and Mazharimousavi, S. Habib},
  title   = {Black holes in double-logarithmic nonlinear electrodynamics},
  journal = {Phys. Scr.},
  volume  = {96},
  pages   = {095213},
  year    = {2021},
  doi     = {10.1088/1402-4896/ac098f},
  url     = {https://doi.org/10.1088/1402-4896/ac098f}
}

@article{Mazharimousavi2024,
    author = {Mazharimousavi, S. Habib},
    title = {Confinement and nonlinear electrodynamics: asymptotic Schwarzschild charged black hole},
    journal = {Phys. Dark Univ. },
    volume = {43},
    pages = {101413},
    year = {2024},
    doi = {10.1016/j.dark.2023.101413},
    url = {https://doi.org/10.1016/j.dark.2023.101413}
}

@article{AlBadawiAhmed2025,
    author = {Al-Badawi, Ahmad and Ahmed, Faizuddin},
    title = {A new black hole coupled with nonlinear electrodynamics surrounded by quintessence: Thermodynamics, geodesics, and Regge--Wheeler potential},
    journal = {Chin. J Phys.},
    volume = {94},
    pages = {185--203},
    year = {2025},
    doi = {10.1016/j.cjph.2025.01.021},
    url = {https://doi.org/10.1016/j.cjph.2025.01.021}
}

@article{EulerHeisenberg1936,
    author = {Heisenberg, W. and Euler, H.},
    title = {Folgerungen aus der Diracschen Theorie des Positrons},
    journal = {Z. Phys.},
    volume = {98},
    pages = {714--732},
    year = {1936},
    doi = {10.1007/BF01343663}
}

@article{Schwinger1951,
    author = {Schwinger, J.},
    title = {On Gauge Invariance and Vacuum Polarization},
    journal = {Phys. Rev.},
    volume = {82},
    pages = {664--679},
    year = {1951},
    doi = {10.1103/PhysRev.82.664}
}

@article{Soleng1995,
    author = {Soleng, H. H.},
    title = {Charged black points in general relativity coupled to the logarithmic U(1) gauge theory},
    journal = {Phys. Rev. D},
    volume = {52},
    pages = {6178--6181},
    year = {1995},
    doi = {10.1103/PhysRevD.52.6178}
}

@article{Nascimento2024,
    author = {Nascimento, Francinaldo Florencio do and Bezerra, Valdir Barbosa and Toledo, Jefferson de Morais},
    title = {Black Holes with a Cloud of Strings and Quintessence in a Non-Linear Electrodynamics Scenario},
    journal = {Universe},
    volume = {10},
    number = {11},
    pages = {430},
    year = {2024},
    doi = {10.3390/universe10110430},
    url = {https://doi.org/10.3390/universe10110430}
}

@article{Hayward2006,
author = {Sean A. Hayward},
title = {Formation and Evaporation of Nonsingular Black Holes},
journal = {Phys. Rev. Lett.},
volume = {96},
pages = {031103},
year = {2006},
doi = {10.1103/PhysRevLett.96.031103}
}

@book{Wald1984,
  author    = {Robert M. Wald},
  title     = {General Relativity},
  publisher = {University of Chicago Press},
  address   = {Chicago, IL},
  year      = {1984}
}

@article{EHTL1,
  author        = {Akiyama, Kazunori and others},
  collaboration = {Event Horizon Telescope},
  title         = {{First M87 Event Horizon Telescope Results.
                   I. The Shadow of the Supermassive Black Hole}},
  journal       = {Astrophys. J. Lett.},
  volume        = {875},
  pages         = {L1},
  year          = {2019},
  doi           = {10.3847/2041-8213/ab0ec7},
}

@article{EHTL6,
  author        = {Akiyama, Kazunori and others},
  collaboration = {Event Horizon Telescope},
  title         = {{First M87 Event Horizon Telescope Results.
                   VI. The Shadow and Mass of the Central Black Hole}},
  journal       = {Astrophys. J. Lett.},
  volume        = {875},
  pages         = {L6},
  year          = {2019},
  doi           = {10.3847/2041-8213/ab1141},
}

@article{EHTL12,
  author        = {Akiyama, Kazunori and others},
  collaboration = {Event Horizon Telescope},
  title         = {{First Sagittarius A* Event Horizon Telescope Results. I. The Shadow of the Supermassive Black Hole
                   in the Center of the Milky Way}},
  journal       = {Astrophys. J. Lett.},
  volume        = {930},
  pages         = {L12},
  year          = {2022},
  doi           = {10.3847/2041-8213/ac6674},
}

@article{EHTL17,
  author        = {Akiyama, Kazunori and others},
  collaboration = {Event Horizon Telescope},
  title         = {{First Sagittarius A* Event Horizon Telescope Results.
                   VI. Testing the Black Hole Metric}},
  journal       = {Astrophys. J. Lett.},
  volume        = {930},
  pages         = {L17},
  year          = {2022},
  doi           = {10.3847/2041-8213/ac6756},
}

@article{Lessa2020,
  author        = {Lessa, L. A. and Silva, J. E. G. and Maluf, R. V. and Almeida, C. A. S.},
  title         = {Modified black hole solution with a background Kalb-Ramond field},
  journal       = {Eur. Phys. J. C},
  volume        = {80},
  pages         = {335},
  year          = {2020},
  doi           = {10.1140/epjc/s10052-020-7902-1},
}

@article{Volker2022,
  author  = {Volker Perlick and Oleg Yu. Tsupko},
  title   = {Calculating black hole shadows: Review of analytical studies},
  journal = {Phys. Rep.},
  volume  = {947},
  pages   = {1--39},
  year    = {2022},
  doi     = {10.1016/j.physrep.2021.10.004},
  url     = {https://doi.org/10.1016/j.physrep.2021.10.004}
}

@article{KalbRamond1974,
    author  = {Kalb, Michael and Ramond, Pierre},
    title   = {{Classical direct interstring action}},
    journal = {Phys. Rev. D},
    volume  = {9},
    pages   = {2273},
    year    = {1974},
    doi     = {10.1103/PhysRevD.9.2273}
}

@article{Yang2023,
  author        = {Yang, Ke and Chen, Yi-Zhong and Duan, Zhao-Qiang and Zhao, Ji-Yuan},
  title         = {Static and spherically symmetric black holes in gravity with a background Kalb-Ramond field},
  journal       = {Phys. Rev. D},
  volume        = {108},
  pages         = {124004},
  year          = {2023},
  doi           = {10.1103/PhysRevD.108.124004},
}

@article{Belchior2025GlobalMonopole,
  author  = {Belchior, F. M. and Maluf, R. V. and Petrov, A. Y. and others},
  title   = {Global monopole in a Ricci-coupled Kalb--Ramond bumblebee gravity},
  journal = {Eur. Phys. J C},
  volume  = {85},
  pages   = {658},
  year    = {2025},
  doi     = {10.1140/epjc/s10052-025-14390-9}
}

@article{Vagnozzi2023,
  author        = {Vagnozzi, Sunny and others},
  title         = {{Horizon-scale tests of gravity theories and fundamental physics from the Event Horizon Telescope image of
                   Sagittarius~A$^*$}},
  journal       = {Class. Quant. Grav.},
  volume        = {40},
  pages         = {165007},
  year          = {2023},
  doi           = {10.1088/1361-6382/acd97b},
}

@article{Duan2024,
    author = {Duan, Z. Q. and Zhao, J. Y. and Yang, K.},
    title = {Electrically charged black holes in gravity with a background Kalb--Ramond field},
    journal = {Eur. Phys. J C},
    volume = {84},
    pages = {798},
    year = {2024},
    doi = {10.1140/epjc/s10052-024-13188-5}
}

@article{Kruglov2022,
  author        = {Kruglov, Sergey I.},
  title         = {{Magnetic black holes with generalized ModMax model of nonlinear electrodynamics}},
  journal       = {Int. J. Mod. Phys. D},
  volume        = {31},
  pages         = {2250025},
  year          = {2022},
  doi           = {10.1142/S0218271822500250},
}

@article{Ahmed2026a,
  author       = {Faizuddin Ahmed and Ahmad Al-Badawi and Izzet Sakalli},
  title        = {Constraining Kalb-Ramond gravity with cloud of strings using EHT shadow observations and X-ray binary QPO data},
  journal      = {Phys. Dark Univ.},
  volume       = {52},
  pages        = {102315},
  year         = {2026},
  doi          = {https://doi.org/10.1016/j.dark.2026.102315},
}

@article{AhmedBadawiSakalli2026,
    author = {F. Ahmed and A. Al-Badawi and I. Sakall{\i}},
    title = {Geodesic structure, perturbative dynamics and thermal properties of black hole in Kalb–Ramond gravity},
    journal = {Mod. Phys. Lett. A},
    volume = {41},
    pages = {2650061},
    year = {2026},
    doi = {10.1142/S0217732326500616}
}

@article{AhmedBadawiSakalli2025,
  author  = {A. Al-Badawi and F. Ahmed and I. Sakall{\i}},
  title   = {Particle dynamics and thermal properties in Kalb--Ramond ModMax black holes: Theoretical predictions for observational tests of exotic physics},
  journal = {Phys. Dark Univ.},
  volume  = {50},
  pages   = {102076},
  year    = {2025},
  doi     = {10.1016/j.dark.2025.102076}
}

\end{document}